\documentclass[12pt]{article}
\RequirePackage{algorithm,algorithmic,amssymb,epsf,epic}

\def\denseformat{
\setlength{\textheight}{9in}
\setlength{\textwidth}{6.9in}
\setlength{\evensidemargin}{-0.2in}
\setlength{\oddsidemargin}{-0.2in}
\setlength{\headsep}{10pt}
\setlength{\topmargin}{-0.3in}
\setlength{\columnsep}{0.375in}
\setlength{\itemsep}{0pt}
}

\newtheorem{theorem}{Theorem}[section]

\newtheorem{definition}[theorem]{Definition}

\newtheorem{lemma}[theorem]{Lemma}

\newtheorem{corollary}[theorem]{Corollary}

\newtheorem{comment}[theorem]{Comment}

\def\boldhead#1:{\par\vskip 7pt\noindent{\bf #1:}\hskip 10pt}
\def\ithead#1:{\par\vskip 7pt\noindent{\it #1:}\hskip 10pt}

\def\inline#1:{\par\vskip 7pt\noindent{\bf #1:}\hskip 10pt}
\def\midinline#1:{\par\noindent{\bf #1:}\hskip 10pt}
\def\dnsinline#1:{\par\vskip -7pt\noindent{\bf #1:}\hskip 10pt}
\def\ddnsinline#1:{\newline{\bf #1:}\hskip 10pt}
\def\largeinline#1:{\par\vskip 7pt\noindent{\large\bf #1:}\hskip 10pt}
\long\def\comment #1\commentend{}
\long\def\commhide #1\commhideend{}
\long\def\commfull #1\commend{#1}
\long\def\commabs #1\commenda{}
\long\def\commtim #1\commendt{#1}
\long\def\commb #1\commbend{}
\long\def\commedit #1\commeditend{} 

\long\def\commB #1\commBend{}       

\long\def\commex #1\commexend{}     

\long\def\commsiena #1\commsienaend{}  
                                         
\long\def\commBI #1\commBIend{}  
                                         
\long\def\CProof #1\CQED{}

\def\blackslug{\hbox{\hskip 1pt \vrule width 4pt height 8pt
    depth 1.5pt \hskip 1pt}}
\def\QED{\quad\blackslug\lower 8.5pt\null\par}

\long\def\PPP#1{\noindent{\bf Proof:}{ #1}{\quad\blackslug\lower 8.5pt\null}}

\long\def\denspar #1\densend
{#1}

\newif\ifnotesw\noteswtrue
   {\ifnotesw\marginpar[\hfill\(\top\)]{\(\top\)}\fi}%
   {\ifnotesw\marginpar[\hfill\(\bot\)]{\(\bot\)}\fi}

\newcommand{\mnote}[1]%
    {\ifnotesw\marginpar%
        [{\scriptsize\it\begin{minipage}[t]{\marginparwidth}
        \raggedleft#1%
                        \end{minipage}}]%
        {\scriptsize\it\begin{minipage}[t]{\marginparwidth}
        \raggedright#1%
                        \end{minipage}}%
    \fi}

\def\cB{{\cal B}}

\def\cD{{\cal D}}

\def\cH{{\cal H}}

\def\cJ{{\cal J}}

\def\cL{{\cal L}}

\def\cP{{\cal P}}

\def\cS{{\cal S}}

\def\tO{{\tilde O}}

\def\MathF{\hbox{\rm I\kern-2pt F}}
\def\MathP{\hbox{\rm I\kern-2pt P}}
\def\MathR{\hbox{\rm I\kern-2pt R}}
\def\MathZ{\hbox{\sf Z\kern-4pt Z}}
\def\MathN{\hbox{\rm I\kern-2pt I\kern-3.1pt N}}
\def\MathC{\hbox{\rm \kern0.7pt\raise0.8pt\hbox{\footnotesize I}
\kern-4.2pt C}}
\def\MathQ{\hbox{\rm I\kern-6pt Q}}

\def\MathE{\hbox{{\rm I}\hskip -2pt {\rm E}}} 

\newsavebox{\ttop}\newsavebox{\bbot}

\def\eps{\epsilon}

\def\setmns{\setminus}

\def\nin{{~\not \in~}}
\def\emset{\emptyset}

\newcommand{\Prob}{\MathP}
\newcommand{\Expect}{\MathE}

\def\etal{\emph{et~al.}}

\denseformat

\begin{document}

\renewcommand{\d}[1]{\ensuremath{\operatorname{d}\!{#1}}}

\def\wmax{{w_{max}}}
\def\tO{\tilde{O}}
\def\nin{\not \in}
\def\emset{\emptyset}
\def\setmns{\setminus}
\def\etal{{et al.~}}
\def\Pairs{\mathrm{Pairs}}
\def\Paths{\mathrm{Paths}}
\def\pred{\mathrm{pred}}
\def\succ{\mathrm{succ}}
\def\NULL{\mathrm{NULL}}
\def\exp{\mathrm{exp}}
\def\Ball{\mathrm{Ball}}
\def\tPi{\tilde{\Pi}}
\def\TB{\cB^{{1/3}}}
\def\Branch{\mathrm{Branch}}
\def\uzero{u^{(0)}}
\def\uone{u^{(1)}}
\def\uj{u^{(j)}}
\def\vzero{v^{(0)}}
\def\vone{v^{(1)}}
\def\vj{v^{(j)}}
\def\dzero{d^{(0)}}
\def\done{d^{(1)}}
\def\dj{d^{(j)}}
\def\dG{d_G}
\def\third{{1 \over 3}}
\def\stretchexp{{\log_{4/3} 7}}
\def\tLambda{\tilde{\Lambda}}
\newcommand{\Patrascu}{P\v{a}tra\c{s}cu{~}}

\title{A Linear-Size Logarithmic Stretch \\ Path-Reporting Distance  Oracle for General Graphs\footnote{A preliminary version of this paper was published in SODA'15 \cite{EP15}.}
}
\author{
Michael Elkin
\thanks{Department of Computer Science, Ben-Gurion University of the
  Negev, Beer-Sheva, 84105, Israel, {\tt elkinm@cs.bgu.ac.il}
\newline
This research has been supported by the Israeli Academy of Science,
grant 593/11, and by the Binational Science Foundation, grant 2008390.
In addition, this research has been supported by the Lynn and William Frankel Center for Computer Science.
A part of this research was performed while visiting the  the Center for Massive Algorithms (MADALGO), which is supported by Danish National Research Foundation grant DNRF84.
}
\and
Seth Pettie
\thanks{Department of Computer Science, University of Michigan, Ann Arbor.
\newline
This research has been supported 
by the Binational Science Foundation, grant 2008390.
A part of this research was performed while visiting the Center for Massive Algorithms (MADALGO), which is supported by Danish National Research Foundation grant DNRF84.}
}


\maketitle

\begin{abstract}
In  \cite{TZ01}  Thorup and Zwick came up with a landmark distance oracle. Given an  $n$-vertex undirected graph $G = (V,E)$ and a parameter $k = 1,2,\ldots$, their  oracle has  size $O(k n^{1+1/k})$, and  upon a query $(u,v)$ it  constructs a path $\Pi$ between $u$ and $v$ of length $\delta(u,v)$ such that $d_G(u,v) \le \delta(u,v) \le (2k-1) d_G(u,v)$.
The query time of the oracle from \cite{TZ01} is $O(k)$ (in addition to the length of the returned path), and it was subsequently improved to $O(1)$ \cite{WN12,C14}.
A major drawback of the oracle of \cite{TZ01} is that its space is $\Omega(n \cdot \log n)$. Mendel and Naor \cite{MN06} devised an oracle with space $O(n^{1+1/k})$ and stretch $O(k)$, but their oracle can only report distance estimates and not actual paths. In this paper we devise a path-reporting distance oracle with size $O(n^{1+1/k})$, 
stretch $O(k)$ and query time $O(n^\eps)$, for an arbitrarily small $\eps > 0$. In particular, for $k = \log n$ our oracle provides logarithmic stretch using linear size.
Another variant of our oracle has size $O(n \log\log n)$, polylogarithmic stretch, and query time $O(\log\log n)$.

For unweighted graphs we devise a distance oracle with multiplicative stretch $O(1)$, additive stretch $O(\beta(k))$, for a function $\beta(\cdot)$, space $O(n^{1+1/k} \cdot \beta)$, and query time $O(n^\eps)$, for an arbitrarily small constant $\eps >0$. The tradeoff between multiplicative stretch and size in these oracles is far below Erd\H{o}s's girth conjecture threshold (which is stretch $2k-1$ and size $O(n^{1+1/k})$). Breaking the girth conjecture tradeoff is achieved by exhibiting a tradeoff of different nature between additive stretch $\beta(k)$ and size $O(n^{1+1/k})$. A similar type of tradeoff was exhibited by a construction of $(1+\eps,\beta)$-spanners due to Elkin and Peleg \cite{EP01}. However, so far $(1+\eps,\beta)$-spanners had no counterpart in the distance oracles' world.

An important novel tool that we develop on the way to these results is a {distance-preserving path-reporting oracle}.  We believe that this oracle is of independent interest.

\end{abstract}



\section{Introduction}

\subsection{Distance Oracles for General Graphs}

In the {\em distance oracle} problem we wish to preprocess a weighted undirected $n$-vertex graph $G = (V,E)$. As a result of this preprocessing we construct a compact data structure (which is called {\em distance oracle}) $\cD(G)$, which given a query pair $(u,v)$ of vertices will efficiently return a distance estimate $\delta(u,v)$ of the distance $\dG(u,v)$ between $u$ and $v$ in $G$. Moreover, the distance oracle should also compute an actual path $\Pi(u,v)$ of length $\delta(u,v)$  between these vertices in $G$. We say that a distance oracle is {\em path-reporting} if it does produce the paths $\Pi(u,v)$ as above; otherwise we say that it is not path-reporting.

The most important parameters of a distance oracle are its stretch, its size, and its worst-case query time.\footnote{The query time of all path-reporting distance oracles that we will discuss is of the form $O(q + |\Pi|)$, where $\Pi$ is the path returned by the query algorithm. To simplify the notation we will often omit the additive term of $O(|\Pi|)$.}
 The {\em stretch} $\alpha$ of a distance oracle $\cD(G)$ is the smallest (in fact, infimum) value such that for every $u,v \in V$,
 $d_G(u,v) \le \delta(u,v) \le \alpha \cdot d_G(u,v)$. 

The term {\em distance oracle} was coined by Thorup and Zwick \cite{TZ01}. See their paper also for a very persuasive motivation of this natural notion. In their seminal paper Thorup and Zwick \cite{TZ01} devised a path-reporting distance oracle (henceforth, TZ oracle). The TZ oracle with a  parameter $k =1,2,\ldots$ has size $O(k \cdot n^{1+1/k})$, stretch $2k-1$ and query time $O(k)$. As argued in \cite{TZ01}, this tradeoff between size and stretch is essentially optimal for $k \le {{\log n} \over {\log\log n}}$, as Erdos' girth conjecture implies that $\Omega(n^{1+1/k})$ space is required for any $k$. Note, however, that $k \cdot n^{1+1/k} = \Omega(n \cdot \log n)$, and Thorup and Zwick \cite{TZ01} left it open if one can obtain meaningful distance oracles of {\em linear} size (or, more generally, size $o(n \log n)$). 

A partial answer to this  question was provided by Mendel and Naor \cite{MN06}, who  devised a distance oracle with size $O(n^{1+1/k})$, stretch $O(k)$ and query time $O(1)$. Alas, their distance oracle is {\em inherently not path-reporting}.  Specifically, the oracle of \cite{MN06} stores a collection of $O(k \cdot n^{1/k})$ hierarchically-separated trees (henceforth, HSTs; see \cite{Bar96} for its definition), whose sizes sum up to $O(n^{1+1/k})$.
The query algorithm for this oracle can return paths from these HSTs, i.e., paths which at best can belong to the metric closure of the original graph. These paths will typically not belong to the graph itself.

 One can try to convert this collection into a collection of low-stretch spanning trees of the input graph $G$ using star-decomposition or petal-decomposition techniques (see \cite{EEST05,AN12}). However, each of this spanning trees is doomed to have $n-1$ edges, making the size of the entire structure as large as $\Omega(k \cdot n^{1+1/k})$. (In addition, with the current state-of-the-art techniques with low-stretch spanning trees one can only achieve bounds which are somewhat worse than the optimal ones achievable with HSTs. Hence the approach that we have just outlined will probably produce an oracle with stretch $\omega(k)$, while using space $O(k \cdot n^{1+1/k})$.)

Another result in this direction was recently obtained by Elkin, Neiman and Wulff-Nilsen \cite{ENWN14}. For a parameter $t \ge 1$ their oracle uses space $O(n \cdot t)$ and provides stretch $O(\sqrt{t} \cdot n^{2/\sqrt{t}})$ for weighted graphs. The query time of their oracle is $O(\log t \cdot \log_n \wmax)$, where $\wmax$ is the aspect ratio of the graph, i.e., the ratio between the heaviest and the lightest edge.
For unweighted graphs their oracle exhibits roughly the same behavior. For a parameter $\eps >0$ it uses space $O(n \cdot t/\eps)$ and provides stretch $O(t \cdot n^{1/t}(t + n^{\eps/t}))$. 

The distance oracles of \cite{ENWN14} are the first path-reporting oracles that use $o(n \log n)$ space and provide non-trivial stretch. However, their stretch is by far larger than that of the oracles of \cite{TZ01,MN06}. Therefore the tantalizing problem of whether one can have a linear-size path-reporting distance oracle with logarithmic stretch remained wide open. In the 
current paper we answer this question in the affirmative. For any $k$, ${{\log n} \over {\log\log n}} \le k \le \log n$, and any arbitrarily small constant $\eps > 0$, our path-reporting distance oracle has stretch $O(k)$, size $O(n^{1+1/k})$ and query time $O(n^\eps)$. (When $\eps >0$ is subconstant the stretch becomes $O(k)\cdot (1/\eps)^{O(1)}$.) Hence  our oracle achieves an optimal up to constant factors tradeoff between size and stretch in the  range 
${{\log n} \over {\log\log n}} \le k \le \log n$, i.e., in the range "missing" in the Thorup-Zwick's result. Though our query time is $n^\eps$ for an arbitrarily small constant  $\eps >0$ is much larger than Thorup-Zwick's query time, we stress that all existing path-reporting distance oracles either use space $\Omega(n \cdot \log n)$ \cite{TZ01,WN12,C14} or have stretch $n^{\Omega(1)}$ \cite{ENWN14}. (The query time of the TZ oracle was recently improved to $O(1)$ in \cite{WN12,C14}.) The only previously existing path-reporting distance oracle that achieves the optimal tradeoff in this range of parameters can be obtained by constructing a $(2k-1)$-spanner\footnote{For a parameter $t \ge 1$,  $G' = (V,H)$   is a {\em $t$-spanner} of a graph $G = (V,E)$, $H \subseteq E$, if $d_H(u,v) \le t \cdot d_G(u,v)$.} with $O(n^{1+1/k})$ edges and answering queries by conducting Dijkstra explorations in the spanner. However, with this  approach the query time is 
 $O(n^{1+1/k})$. Our result is a drastic improvement of this trivial bound from $O(n^{1+1/k})$ to $O(n^\eps)$, for an arbitrarily small constant $\eps > 0$. 

We also can trade between the stretch and the query time. Specifically, a variant of our oracle uses $O(n \log\log n)$ space, has stretch $O(\log^{\stretchexp} n) \approx O(\log^{6.76} n)$ and query time $O(\log\log n)$. For a comparison, the path-reporting distance oracle of \cite{ENWN14} with this stretch uses space
 $\Omega(n \cdot {{\log n} \over {\log\log n}})$ and has query time $O(\log\log n \cdot \log_n \wmax)$. 

\comment
We believe that pushing the size bound from $O(n\log n)$ down to $O(n)$ in the context  of path-reporting distance oracles for general graphs is a natural and fundamental problem. In fact, Kawabarayashi \etal \cite{KKS11} did this in the context of minor-free graphs. They argue that in many practical scenarios the overhead of $\Theta(\log n)$ is intolerable. Also, in a closely related setting of $(1+\eps,\beta)$-spanners (see Section \ref{sec:unw}) there was a lot of research effort invested in pushing the size bound down from $O(n) \cdot (\log n)^{O(\log\log\log  n)}$ (due to \cite{EP01}) to $O(n \cdot \log\log n)$ \cite{Pet09}.
\commentend

We also remark that using a super-constant (but not trivial) query time is a common place by now in the distance oracles literature. In particular, this is the case in the oracles of Porat and Roditty \cite{PR13}, Agarwal and Godfrey \cite{AG13}  and of Agarwal \etal \cite{AGHP11}.

\subsection{Distance Oracles with Stretch $(\alpha,\beta)$ for Unweighted Graphs}
\label{sec:unw}

We say that a distance oracle $\cD(G)$ provides stretch $(\alpha,\beta)$ for a pair of parameters $\alpha \ge 1, \beta \ge 0$ if for any query $(u,v)$ it constructs a path $\Pi(u,v)$ of length $\delta(u,v)$ which satisfies $d_G(u,v) \le \delta(u,v) \le \alpha \cdot d_G(u,v) + \beta$. The notion of $(\alpha,\beta)$-stretch is originated from the closely related area of {\em spanners}. A subgraph $G' = (V,H)$ is said to be an {\em $(\alpha,\beta)$-spanner} of a graph $G = (V,E)$ , $H \subseteq E$, if for every pair $u,v \in V$, it holds that $d_H(u,v) \le \alpha \cdot d_G(u,v) + \beta$.

This notion was introduced in \cite{EP01}, where it was shown that for any $\eps > 0$ and $k = 1,2,\ldots$, for any $n$-vertex unweighted graph $G = (V,E)$ there exists a $(1+\eps,\beta)$-spanner  with $O(\beta \cdot n^{1+1/k})$ edges, where $\beta= \beta(\eps,k)$ is independent of $n$. Later a number of additional constructions of $(1+\eps,\beta)$-spanners with similar properties were devised in \cite{E01,TZ06,Pet09}.

It is natural to attempt converting these constructions of  spanners into distance oracles with a similar tradeoff between stretch and size. However, generally so far such attempts were not successful. See, e.g., the discussion titled "Additive Guarantees in Distance Oracles"  in the introduction of \cite{PR10}. \Patrascu and Roditty \cite{PR10} devised a distance oracle with stretch $(2,1) $ and size $O(n^{5/3})$, and query time $O(1)$. Abraham and Gavoille \cite{AG11} generalized the result of \cite{PR10} to devise a distance oracle with stretch $(2k-2,1)$ and space $\tO(n^{1 + (2/(2k-1))})$. (The query time in \cite{AG11} is unspecified.) 

Note, however, that neither of these previous results  achieves multiplicative stretch $o(k)$ with size $O(n^{1+1/k})$, at the expense of an additive stretch. (This is the case with the result of \cite{EP01} in the context of {\em spanners}, where the multiplicative stretch becomes as small as $1 + \eps$, for an arbitrarily small $\eps > 0$.)
In this paper we devise the first distance oracles that do achieve such a tradeoff. Specifically, our path-reporting distance oracle has stretch $(O(1),\beta(k))$, space $O(\beta(k) \cdot n^{1+1/k})$, $\beta(k) = k^{O(\log\log k)}$, and query time $O(n^\eps)$, for an arbitrarily small $\eps > 0$. 
The multiplicative stretch $O(1)$ here is a polynomial function of $1/\eps$, but it can be made much smaller than $k$. (Think, e.g., of $\eps > 0$ being a constant  and $k$ being a slowly growing function of $n$.) We can also have stretch $(o(k),\beta(k))$, space $O(\beta(k) \cdot n^{1+1/k})$ and query time $n^{O(k^{-\gamma})}$, where $\gamma > 0$ is a universal constant. (Specifically, the theorem holds, e.g.,  for $\gamma = 1/7$.)

In both these results the tradeoff between  multiplicative stretch and size of the oracle is below Erd\H{o}s' girth conjecture barrier (which is stretch $2k-1$ and space $O(n^{1+1/k})$). 
In fact, it is known that when the additive stretch is 0, distance oracles for general $n$-vertex graphs that have size $O(n^{1+1/k})$ must have multiplicative stretch $\Omega(k)$ \cite{TZ01,LPS88,LU95}.
Our results, like the results of \cite{EP01} for spanners, break this barrier by introducing an additive stretch $\beta(k)$. To the best of our knowledge, our distance oracles are the first distance oracles that exhibit this behavior.

Using known lower bounds we also show that there exist no distance labeling schemes with stretch $(O(1),\beta(k))$ and maximum label size $O(\beta(k) \cdot n^{1/k})$.
(Rather one needs labels of size $n^{\Omega(1)}$ for this.)
This is also the case for routing schemes. (See Section \ref{sec:prel} for relevant definitions.) We also show that in the cell-probe model of computation any distance oracle for unweighted undirected $n$-vertex graphs with stretch $(O(1),\beta(k))$ and space $O(\beta(k) \cdot n^{1+1/k})$ has query time $\Omega(k)$. This is in contrast to distance oracles with multiplicative stretch, which can have constant query time \cite{MN06,C14}.

\subsection{Distance Oracles for Sparse Graphs}
\label{sec:intr_sparse}

A central ingredient in all our distance oracles is a new path-reporting distance oracle for graphs with $O(n)$ edges. The most relevant result in  this context is the paper by Agarwal \etal \cite{AGHP11}. In this paper the authors devised a (not path-reporting)\footnote{It was erroneously claimed in \cite{AGHP11} that all their distance oracles are path-reporting. While  their distance oracles with stretch smaller than 3 are path-reporting (albeit their space requirement is superlinear), this is not the case for their oracles with stretch $4k-1$, $k \ge 1$ \cite{Agar_priv}.} linear-size distance oracle which given a parameter $k=1,2,\ldots$ provides distance estimates with stretch $4k-1$, uses linear space and has time $O(n^{1/(k+1)})$.  (Their result is, in fact, more general than this. We provide this form of their result to facilitate the comparison.)
In this paper we present the first path-reporting linear-size distance oracle for this range of parameters. Specifically, our linear-size oracle (see Corollary \ref{cor:multilevel}) has stretch $O(k^\stretchexp)$ and query time $O(n^{1/k})$, for any constant parameter $k$ of the form $k = (4/3)^h$, $h=1,2,\ldots$.

\subsection{A Distance-Preserving Path-Reporting Distance Oracle}

In \cite{CE05} the authors showed that for any $n$-vertex graph $G = (V,E)$ and a collection $\cP$ of $P$ pairs of vertices there exists a subgraph $G' = (V,H)$ of size $O(\max\{n + \sqrt{n}\cdot P,\sqrt{P} \cdot n\})$ so that for every $(u,v) \in \cP$, $d_H(u,v) = d_G(u,v)$.
In this paper we devise the first distance-oracle counterpart of this result. Specifically, our distance oracle uses $O(n + P^2)$ space, and for any query $(u,v) \in \cP$ it produces the exact shortest path $\Pi$ between $u$ and $v$ in $O(|\Pi|)$ time, where $|\Pi|$ is the number of edges in $\Pi$.

We employ this distance oracle very heavily in all our other constructions.

\inline Remark:
The construction time of our distance-preserving oracle is $O(n \cdot P^2) + \tilde{O}(m \cdot \min\{n,P\})$.
The construction time of our path-reporting oracle for sparse graphs is $\tO(m \cdot n) = \tO(n^2 \lambda)$, where $\lambda = m/n$.
The construction time of our  oracles with nearly-linear space for general graphs is $\tO(n^{2+1/k})$.
Finally, the construction time of our oracle for unweighted graphs with a hybrid multiplicative-additive stretch is $\tO(\beta(k) n^{2+1/k}) = k^{O(\log\log k)} \tO(n^{2+1/k})$. (In both cases  
$k$ is  the stretch parameter of the respective oracle.)

\subsection{Related Work}

There is a huge body of literature about distance oracles by now. In addition to what we have already surveyed there are probe-complexity lower bounds by Sommer \etal \cite{SVY09}. There is an important line of work by \Patrascu \etal \cite{PRT12,PR10} on oracles with rational stretch. 
Finally, Baswana and Sen \cite{BS06}, Baswana and Kavitha \cite{BK06} and
Baswana \etal \cite{BGSU08} improved the preprocessing time of the TZ oracle.

\subsection{Structure of the Paper}

We start with describing our distance preserving oracle (Section \ref{sec:dppro}). We then proceed with devising our basic path-reporting oracle for sparse graphs (Section \ref{sec:basic}).
This oracle can be viewed as a composition of an oracle from Agarwal \etal \cite{AGHP11} with our distance-preserving oracle from Section \ref{sec:dppro}.
The oracle is described for graphs with small arboricity. Its extension to general sparse graphs (based on a reduction from \cite{AGHP11}) is  described in Section \ref{sec:extension}. Then we devise a much more elaborate multi-level path-reporting oracle for sparse graphs. The oracle of \cite{AGHP11} and  our basic oracle from Section \ref{sec:basic} both use just one set of sampled vertices. Our multi-level oracle uses a carefully constructed hierarchy of sampled sets which enables us to get the query time down  from $n^{1/2 +\eps}$ to $n^\eps$. Next we proceed (Section \ref{sec:faster_oracles}) to using this multi-level oracle for a number of applications.
Specifically, we use it to construct a linear-size logarithmic stretch path-reporting oracle with query time $n^\eps$,  linear-size polylogarithmic stretch path-reporting oracle with query time $O(\log\log n)$, and finally,  oracles that break the girth barrier  for unweighted graphs.
Our lower bounds can be found in Section \ref{sec:lb}.

\section{Preliminaries}
\label{sec:prel}

For a pair of integers $a \le b$, we denote $[a,b] = \{a,a+1,\ldots,b\}$, and $[b] = [1,b]$. The arboricity of a graph $G$ is given by $\lambda(G) = \max_{U \subseteq V, |U| \ge 2} {{|E(U)|} \over {|U|-1}}$, where $E(U)$ is the set of edges induced by the vertex set $U$.
We denote by $\deg_G(u)$ the degree of a vertex $u$ in $G$; we omit $G$ from this notation whenever $G$ can be understood from the context.
We use the notation $\tO(f(n)) = O(f(n) \mathrm{polylog}(f(n)))$ and $\tilde{\Omega}(f(n) = \Omega(f(n)/\mathrm{polylog}(f(n)))$.
We say that a function $f()$ is {\em quasi-polynomial} if $f(n) \le n^{\log^{O(1)} n}$.

A {\em distance-labeling scheme} for a graph $G = (V,E)$ assigns every vertex $v \in V$ a short label $\varphi(v)$. Given a pair of labels $\varphi(u),\varphi(v)$ of a pair of vertices $u,v \in V$, the scheme computes an estimate $\delta(\varphi(u),\varphi(v))$.  This estimate has to be within a factor $\alpha$, for some $\alpha \ge 1$, from the actual distance $d_G(u,v)$ between $u$ and $v$ in $G$. The parameter $\alpha$ is called the {\em stretch} of the labeling scheme, and the maximum number of bits employed by one of the labels is called the {\em (maximum) label size} of the scheme.

A closely related notion is that of {\em compact routing scheme}. Here each vertex $v$ is assigned a label $\varphi(v)$ and a routing table $\psi(v)$. Given a label $\varphi(u)$ of routing destination $u$ and its own routing table $\psi(v)$, the vertex $v = v_0$ needs to be able to compute the next hop $v_1$. Given the table $\psi(v_1)$ of $v_1$ and the destination's label $\varphi(u)$, the vertex $v_1$ computes the next hop $v_2$, etc. The resulting path $v = v_0,v_1,v_2,\ldots$ has to end up eventually in $u$, and its length needs to be at most $\alpha$ times longer than the length of the shortest $u-v$ path in $G$, for a stretch parameter $\alpha \ge 1$. 
In addition to stretch, another important parameter in this context is the maximum number of bits used by the label and the routing table (together) of any individual vertex.
This parameter will be referred to as {\em maximum memory requirement} of a routing scheme.

\section{A Distance-Preserving Path-Reporting Oracle}
\label{sec:dppro}

Consider  a directed weighted $n$-vertex graph $G = (V,E,\omega)$. (The result given in this section applies to both directed and undirected graphs. However, our other distance oracles apply only to undirected graphs.) 
Let $\Pairs \subseteq {V \choose 2}$ be a subset of ordered pairs of vertices. We denote its cardinality by $P = |\Pairs|$. In this section we describe a distance oracle which given a pair $(u,v) \in \Pairs$ returns a shortest path $\Pi_{u,v}$ from $u$ to $v$ in $G$.
The query time of the oracle is proportional to the number of edges (hops) $|\Pi_{u,v}|$ in $\Pi_{u,v}$. The oracle uses $O(n + P^2)$ space.

The construction of the oracle starts with computing a set $\Paths = \{ \Pi_{u,v} \mid (u,v) \in \Pairs\}$ of  shortest paths between pairs of vertices from $\Pairs$. This collection of shortest paths is required to satisfy the  property that if two distinct paths $\Pi,\Pi' \in \Paths$ traverse two common vertices $x$ and $y$ in the same order (i.e., e.g., both traverse first $x$ and then $y$), then they necessarily share the entire subpath between $x$ and $y$. It is argued in \cite{CE05} that this property can be easily achieved.

We will need the following definitions from \cite{CE05}.

For a path $\Pi = (u_0,u_1,\ldots,u_h)$ and a vertex $u_i \in  V(\Pi)$, the {\em  predecessor} of $u_i$ in $\Pi$, denoted $\pred_\Pi(u_i)$, is the vertex $u_{i-1}$
(assuming that $i \ge 1$; otherwise it is defined as $\NULL$), and the {\em successor} of $u_i$ in $\Pi$, denoted $\succ_\Pi(u_i)$, is the vertex $u_{i+1}$ (again, assuming that $i \le h-1$; otherwise it is $\NULL$).

\begin{definition} \cite{CE05}
A {\em branching event} $(\Pi,\Pi',x)$ is a triple with $\Pi,\Pi' \in \Paths$ being two distinct paths and $x \in V(\Pi) \cap V(\Pi')$ be a vertex that belongs to both paths and such that $\{\pred_\Pi(x),\succ_\Pi(x)\} \neq \{\pred_{\Pi'}(x),\succ_{\Pi'}(x)\}$. We will also say that the two paths $\Pi,\Pi'$ branch at the vertex $x$.
\end{definition}

Note that under this definition if $\Pi$ traverses edges $(u_{i-1},u_i), (u_i,u_{i+1})$ and $\Pi'$ traverses edges $(u_{i+1},u_i),(u_i,u_{i-1})$ then $(\Pi,\Pi',u_i)$ is {\em not} a branching event.

It follows directly from the above property of the collection $\Paths$ (see also \cite{CE05}, Lemma 7.5, for a more elaborate discussion) that for every pair of distinct paths $\Pi,\Pi' \in \Paths$, there are at most two branching events that involve that pair of paths.
Let ${\cal B}$ denote the set of branching events.
 The overall number of branching events for the set
 $\Paths$ is $|{\cal B}| \le |Paths|^2 = P^2$. Our oracle will keep $O(1)$ data for each vertex, $O(1)$ data for each branching event, and $O(1)$ data for each path. Hence the oracle stores $O(n + |\cB| + P)$ data in total.  

Specifically, in our oracle for every vertex $v \in V$ we keep an identity of some path $\Pi \in \Paths$ that contains $v$ as an internal point, and two edges of $\Pi$ incident on $v$.
(If there is no path $\Pi \in \Paths$ that contains $v$ as an internal point, then our oracle stores nothing for $v$ in this data structure.) The path $\Pi$ stored for $v$ will be referred to as the {\em home path} of $v$. 

In addition, for every branching event $(\Pi,\Pi',v)$ we keep the (at most four) edges of $\Pi$ and $\Pi'$ incident on $v$. Finally, for every pair $(x,y) \in \Pairs$ we also store the first and the last edges of the path $\Pi_{x,y}$. Observe that the resulting space requirement is at most $O(n + |\cB| + P) = O(n + P^2)$. 
We assume that the branching events are stored in a hash table of linear size, which allows membership queries in $O(1)$ time per query. 

The query algorithm proceeds as follows. Given a pair $(x,y) \in \Pairs$, we find the first edge $(x,x')$ of the path $\Pi_{x,y}$, and "move" to $x'$.
Then we check if $(x',y)$ is the last edge of $\Pi_{x,y}$. If it is then we are done. Otherwise let $\Pi(x')$ denote the home path of $x'$. (Observe that since the vertex $x'$ is an internal vertex in $\Pi_{x,y}$, it follows that there exists a home path $\Pi(x')$ for $x'$.)

Next, we check if $\Pi(x') = \Pi_{x,y}$. (This test is performed by comparing the identities of the two paths.) If it is the case then we fetch the next edge $(x',x'')$ of $\Pi(x')$, and move to $x''$. Otherwise (if $\Pi(x') \neq \Pi(x,y)$) then we check if the triple $(\Pi(x'),\Pi_{x,y},x')$ is a branching event.
This check is performed by querying the branching events' hash table. 

If there is no branching event  $(\Pi(x'),\Pi_{x,y},x')$ then we again fetch the next edge $(x',x'')$ of $\Pi(x')$, and move to $x''$. (In fact, the algorithm does not need to separate between this case and the case that $\Pi(x') = \Pi_{x,y}$. We distinguished between these cases here for clarity of presentation.) 

Finally, if there is a branching event $(\Pi(x'),\Pi_{x,y},x')$ then we fetch from our data structure all the information associated with this event. In particular, we fetch the next edge $(x',x'')$ of $\Pi_{x,y}$, and move to $x''$.

In all cases the procedure then recurses with $x''$. It is easy to verify that using appropriate hash tables all queries can be implemented in $O(1)$ time per vertex, and in total $O(|\Pi_{x,y}|)$ time. We summarize this section with the following theorem.

\begin{theorem}
\label{thm:dppro}
Given a directed weighted graph $G = (V,E,\omega)$ and a collection $\Pairs \subseteq {V \choose 2}$ of pairs of vertices, our distance-preserving path-reporting  oracle (shortly, {\em DPPRO}) reports shortest paths $\Pi_{x,y}$ for query pairs $(x,y) \in \Pairs$ in $O(|\Pi_{x,y}|)$ time. The oracle employs $O(n + |\cB| + P) = O(n + P^2)$ space, where $\cB$ is the set of branching events for a fixed set of shortest paths between pairs of vertices from $\Pairs$, and $P = |\Pairs|$.
\end{theorem}

\comment
\inline Remark: Here and thereafter when we say that a query time of a path-reporting oracle is $t$, for some $t$, we really mean that it is $t$ plus the number of edges in the reported path.
\commentend

One can construct the shortest paths in $\tilde{O}(m \cdot \min\{P,n\})$ time. Then for each vertex $v$ one keeps the list of paths that traverse $v$. For every such path one keeps the  two edges of this path which are  incident on $v$. In overall $O(n \cdot P^2)$ additional time one can use these lists to create the list of branching events. A hash table with them can be constructed in additional $O(P^2)$ time. Hence the overall construction time of this oracle is $\tilde{O}(m \cdot \min \{P,n\}) + O(n \cdot P^2)$. 

Observe that if one is given a set $S$, $|S| = O(n^{1/4})$, of terminals, then Theorem \ref{thm:dppro} provides a linear-size DPPRO (i.e., $O(1)$ words per vertex {\em on average}) which can report shortest paths between all pairs of terminals. It is well-known that any distance labeling scheme which is guaranteed to return exact distances between all pairs of $n^{1/4}$ terminals must use maximum label size $\Omega(n^{1/4})$ \cite{TZ01}. This is also the case for compact routing schemes \cite{TZ01b}. (In the latter case the lower bound of $\Omega(n^{1/4})$ is on the maximum memory requirement of any individual vertex.)

We remark that our DPPRO here employs $O(n + |\cB| + P)$ space, whereas the underlying distance preserver has $O(n + \sqrt{n \cdot |\cB|})$ edges \cite{CE05}. It is plausible that there exists a DPPRO of size $O(n + \sqrt{n \cdot |\cB|})$. We leave this question open.

\section{A Basic Distance Oracle for Graphs with \\  Bounded Arboricity}
\label{sec:basic}

In this section we describe a basic variant of our path-reporting distance oracle for weighted undirected graphs $G = (V,E,\omega)$ of arboricity $\lambda(G) \le \lambda$, for some parameter $\lambda$. (We will mostly use this oracle for constant or small values of $\lambda$. On the other hand, the result is meaningful for higher values of $\lambda$ as well.)
Our oracle reports paths of stretch $O(k)$, for some positive integer parameter $k$. Unlike the partial oracle from Section \ref{sec:dppro}, the oracle in this section is a full one, i.e., it reports paths for all possible queries $(u,v) \in {V \choose 2}$. This is the case also for all our other oracles, which will be described in consequent sections.
The expected query time of our oracle is $O(n^{1/2 + {1 \over {2k+2}}} \cdot \lambda)$. (Whp\footnote{Here and thereafter we use the shortcut "whp"
for "with high probability". The meaning is that the probability is at least $1 - n^{-c}$, for some constant $c \ge 2$.}, the query time is $O(n^{1/2 + {1 \over {2k+2}}} \cdot \log n \cdot \lambda)$.)
The oracle requires $O(n)$ space, in addition to the space required to store the graph $G$ itself.
Observe that for $\lambda = O(1)$ the query time is $O(n^{1/2 + \eps})$, for an arbitrarily small constant $\eps > 0$, while the stretch is $O({1 \over \eps}) = O(1)$.
In Section \ref{sec:extension} we extend this oracle to general $m$-edge $n$-vertex graphs with $\lambda = {m \over  n}$.

Our basic oracle employs just one level of sampled vertices, which we (following the terminology of \cite{AGHP11}) call {\em landmarks}. Each $v \in V$ is sampled independently at random with probability ${\rho \over n}$, where $\rho$ is a parameter which will be determined in the sequel. Denote by $L$ the set of sampled vertices (landmarks). Note that $\Expect(|L|) = \rho$.

For every vertex $v \in V$ we keep the path $\Pi(v)$ to its closest landmark vertex $\ell(v)$, breaking ties arbitrarily. Denote by $D(v)$ the length $w(\Pi(v))$ of this path.
This is a collection of vertex-disjoint shortest paths trees (shortly, SPTs) $\{T(u) \mid u \in L\}$, where  each $T(u)$ is an SPT rooted at $u$ for the subset $\{ v \mid d_G(u,v) \le d_G(u',v), \forall u' \ne u, u,u' \in L\}$. (Ties are broken arbitrarily.) This collection is a forest, and storing it requires $O(n)$ space. 

The oracle also stores the original graph $G$. For the set of landmarks we compute the complete graph $\cL = (L,{L \choose 2},d_G|L)$. Here $d_G|L$ stands for the metric of $G$ restricted to the point set $L$. (In other words, in the landmarks graph $\cL$, for every pair $u,u' \in L$ of distinct landmarks the weight $\omega_\cL(u,u')$ of the edge $(u,u')$ connecting them is defined by $\omega_\cL(u,u') = d_G(u,u')$.)

Next we invoke Thorup-Zwick's distance oracle \cite{TZ01} with a parameter $k$. (Henceforth we will call it the {\em TZ oracle}.) One can also use here Mendel-Naor's oracle \cite{MN06}, but the resulting tradeoff will be somewhat inferior to the one that is obtained via the TZ oracle.
Denote by $\cH$ the TZ distance oracle for the landmarks graph $\cL$. The oracle requires $O(k \cdot |L|^{1+1/k})$ space, and it provides $(2k-1)$-approximate paths $\Pi_{u,u'}$ in $\cL$ for pairs of landmarks $u,u' \in L$. The query time is $O(k)$ (plus $O(|\Pi_{u,u'}|)$).  Observe that some edges of $\Pi_{u,u'}$ may not belong to the original graph $G$. We note also that by using more recent oracles \cite{C14,WN12} one can have here query time $O(1)$, but this improvement is immaterial for our purposes.

The TZ oracle $\cH$ has a useful property that the union $H = \bigcup \{ \Pi_{u,u'} \mid (u,u') \in {L \choose 2}\}$ of all paths that the oracle returns forms a sparse $(2k-1)$-spanner. Specifically, $\Expect(|H|) = O(k \cdot |L|^{1+1/k})$. (This property holds for Mendel-Naor's oracle as well, but there the stretch of the spanner is $O(k)$, where the constant hidden by the $O$-notation is greater than 2. On the other hand, their space requirement is $O(|L|^{1+1/k})$, rather than $O(k \cdot |L|^{1+1/k})$.)
Fix an oracle $\cH$ as above for $|H| = O(k \cdot |L|^{1+1/k})$. Whp such an $\cH$ will be computed by running the procedure that computes the TZ oracle for $O(\log n)$ times.
We will view the spanner $H$ as a collection of pairs of vertices of our original graph $G$.

Finally, we invoke our distance preserving oracle (shortly, DPPRO) from Section \ref{sec:dppro} on the graph $G$ and set $\Pairs = H$. We will refer to this oracle as $\cD(G,H)$. Its size is, with high probability, $O(n + |H|^2) = O(n + k^2  \cdot |L|^{2+2/k})$. Upon a query $(y,y') \in H$, this oracle returns a shortest path $\Pi_{y,y'}$ between $y$ and $y'$ in $G$ in time $O(|\Pi_{y,y'}|)$. 

Observe that $|L|$ is the sum of identical independent indicator random variables $|L| = \sum_{v \in V} I_v$, where $I_v$ is the indicator random variable of the event $\{v \in L\}$. Hence, by Chernoff's inequality, for any constant $\eps > 0$, 
$$\Prob(|L| > (1 + \eps) \Expect(|L|)) = \Prob(|L| > (1+\eps) \cdot \rho) < \exp({-\Omega(\rho)})~.$$
We will set the parameter $\rho$ to be at least $c \log n$, for a sufficiently large constant $c$. This will ensure that whp $|L| = O(\rho)$, and so
 $|L|^{2+2/k} = O(\rho^{2 + 2/k})$.
Set $\rho$ so that $k^2 \cdot \rho^{2+2/k}  = \Theta(n)$, i.e., $\rho = n^{k \over {2k+2}} \cdot {1 \over {k}}$ . This guarantees that aside from the storage needed for the original graph, the total space used by our oracle 
is $O(n)$.

This completes the construction algorithm of our oracle. Next we describe its query algorithm. We need the following definition.
For a vertex $v \in V$, let $\Ball(v) = \{x \mid d_G(v,x) < d_G(v,\ell(v))\}$ denote the set of all vertices $x$ which are closer to $v$ than the closest landmark vertex $\ell(v)$ to $v$. 

Given a pair $u,v$ of vertices of $G$, our oracle starts with testing if $u \in \Ball(v)$ and if $v \in \Ball(u)$.  To test if $u \in \Ball(v)$ we just conduct a Dijkstra exploration rooted at $v$ in the graph $G$, until we discover either $u$ or $\ell(v)$. (Recall that $G$ is stored in our oracle.) If $u$ is discovered before $\ell(v)$ we conclude that $u \in \Ball(v)$, and return the (exact) shortest path between them. Otherwise we conclude that $u \not \in \Ball(v)$. Analogously, the algorithm tests if $v \in Ball(u)$. 

Henceforth we assume that $u \nin \Ball(v)$ and $v \nin \Ball(u)$, and therefore the two searches returned $u' = \ell(u)$, $v' = \ell(v)$, and the shortest paths
$\Pi(u)$ and $\Pi(v)$ between $u$ and $u'$ and between $v$ and $v'$, respectively. 
(In fact, using the forest of SPTs rooted at landmarks that our oracle stores, the query algorithm can compute shortest paths between $u$ and $u'$ and between $v$ and $v'$ in time proportional to the lengths of these paths.) Observe that $d_G(u',v') \le d_G(u',u) + d_G(u,v) + d_G(v,v')$, and $d_G(u',u),d_G(v,v') \le d_G(u,v)$.
Hence $d_G(u',v') \le 3 \cdot d_G(u,v)$.

Then the query algorithm invokes the query algorithm of the oracle $\cH$ for the landmarks graph $\cL$. The latter algorithm returns a path $\Pi' = (u' = z_0,z_1,\ldots,z_h = v')$ in $\cL$ between $u'$ and $v'$. The length $\omega_\cL(\Pi')$ of this path is at most $(2k-1) \cdot d_G(u',v') \le (6k-3) \cdot d_G(u,v)$.
The time required for this computation is $O(k + h)$, where $|\Pi'| = h$.
For each edge $(z_i,z_{i+1}) \in \Pi'$, $i \in [0,h-1]$, we invoke the query algorithm of the DPPRO $\cD(G,H)$. 
(The edges $(z_i,z_{i+1})$  of  the path $\Pi'$ are typically not edges of the original graph.
$H$ is a $(2k-1)$-spanner of $\cL$ produced by the oracle $\cH$. Observe that $\Pi' \subseteq H$, and so $(z_i,z_{i+1}) \in H$, for every index $i \in [0,h-1]$.) The oracle $\cD(G,H)$ returns a path $\tilde{\Pi}_i$ between $z_i$ and $z_{i+1}$ in $G$ of length $\omega_\cL(z_i,z_{i+1}) = d_G(z_i,z_{i+1})$. Let $\tPi = \tPi_0 \cdot \tPi_1 \cdot \ldots \cdot \tPi_{h-1}$ be the concatenation of these paths. Observe that $\tPi$ is a path in $G$ between $z_0 = u'$ and $z_h = v'$, and 
$$\omega(\tPi) ~=~ \sum_{i=0}^{h-1} \omega(\tPi_i) ~ =~ \sum_{i=0}^{h-1} d_G(z_i,z_{i+1}) ~=~
\sum_{i=0}^{h-1} \omega_\cL(z_i,z_{i+1}) ~=~  \omega_\cL(\Pi') \le (6k-3) \cdot d_G(u,v)~.$$
Finally, the query algorithm returns the concatenated path $\hat{\Pi} = \Pi(u) \cdot \tPi \cdot \Pi(v)$ as the approximate path for the pair $u,v$. 
This completes the description of the query algorithm of our basic oracle.
Observe that
$$ \omega(\hat{\Pi}) = \omega(\Pi(u)) + \omega(\tPi) + \omega(\Pi(v)) \le d_G(u,v) + (6k - 3)\cdot d_G(u,v) + d_G(u,v) = (6k-1) \cdot d_G(u,v)~.$$
Next, we analyze the running time of the query algorithm.
First, consider the step that tests if $v \in \Ball(u)$ and if $u \in \Ball(v)$.
Denote by $X$ the random variable that counts the number of vertices discovered by some fixed Dijkstra exploration originated at $u$ before the landmark $\ell(u)$ is discovered.
We order all graph vertices by their distance from $u$ in a non-decreasing order, i.e., $u = u_0,u_1,\ldots,u_{n-1}$, such that $d_G(u,u_i) \le d_G(u,u_j)$ for $i \le j$.
(This is the order in which the aforementioned Dijkstra exploration originated at $u$ discovers them.)
For an integer value $1 \le t \le n-1$, the probability that $X = t$ is equal to the probability that the vertices $u_0,u_1,\ldots,u_{t-1}$ are all not sampled and the vertex $u_t$ is sampled. Hence $X$ is distributed geometrically with the parameter $p = \rho/n$. Hence
\begin{equation}
\label{eq:geom_expect}
\Expect(X) ~=~ \sum_{t=1}^{n-1} (1 - p)^t \cdot p \cdot t ~ \le~ {1 \over p} ~=~ {n \over \rho}~.
\end{equation}
Also, obviously for any positive constant $c$, $\Prob(X > {n \over \rho} c \ln n) \le (1 - \rho/n)^{(n/\rho) c\ln n} \le n^{-c}$, i.e., whp $X = O({n \over \rho} \log n)$.

Recall that the graph $G$ has arboricity at most $\lambda$, and thus any set of $n' \le n$ vertices induces $O(n' \cdot \lambda)$ edges. Hence Dijkstra algorithm traverses expected $O({n \over \rho} \lambda)$ edges, and whp $O({n \over \rho} \lambda \log n)$ edges. In an unweighted graph such exploration requires time linear in the number of edges, and in weighted\footnote{One subtlety: we have to avoid scanning too many edges with just one endpoint in $\Ball(u)$. We store the edges incident to each vertex $x$ in increasing order of their weights, and relax them in that order when $x$ is scanned. As soon as an edge $(x,y)$ is relaxed such that the tentative distance to $y$ is greater than $d_G(u,\ell(u))$ we can dispense with relaxing the remaining edges. Alternatively, a modification of the sampling rule which we describe in Section \ref{sec:extension} also resolves this issue.} graphs the required time is $O({n \over \rho}(\lambda + \log n))$ in expectation, and $O({n \over \rho} \lambda \cdot \log n)$ whp. (Recall that Dijkstra algorithm that scans a subgraph $(V',E')$ requires time $O(|E'| + |V'| \log |V'|)$.)

The second step of our query algorithm queries the distance oracle $\cH$ for the landmarks graph $\cL$. (The query is $(u',v')$, $u' = \ell(u)$, $v' = \ell(v)$.)
This query returns a path $\Pi'$  between $u'$ and $v'$ in $\cL$ in time $O(|\Pi'| + k)$. Finally, for each of the $h = |\Pi'|$ edges $(z_i,z_{i+1})$, $i = 0,1,\ldots,h-1$ of the path $\Pi'$, the query algorithm invokes our DPPRO $\cD(G,H)$ with the query $(z_i,z_{i+1})$. This oracle returns the shortest path $\tPi_i$ between $z_i$ and $z_{i+1}$ in $G$ within time $O(|\tPi_i|)$. Finally, the algorithm returns the concatenated path $\hat{\Pi} = \Pi(u) \cdot \tPi_0 \cdot \tPi_1 \cdot \ldots \cdot \tPi_{h-1} \cdot \Pi(v)$. The running time required for producing the path $\tPi_0 \cdot \ldots \cdot \tPi_{h-1}$ is $O(\sum_{i=0}^{h-1} |\tPi_i|) = O(|\hat{\Pi}|)$, and $|\Pi'| \le |\hat{\Pi}|$. Hence the overall expected running time  of the algorithm is
 $O({n \over \rho} \cdot \lambda + |\hat{\Pi}|)$ for unweighted graphs, and is $O({n \over \rho} \cdot (\lambda +  \log n) + |\hat{\Pi}|)$ for weighted. (Observe that the additive term of $O(k)$ is dominated by $O({n \over \rho} \cdot \lambda)$. Specifically, we will be using $\rho \le n/\log n$, and
$k \le O(\log n)$.) For the high-probability bounds one needs to multiply the first term of the running time by an additional $O(\log n)$ factor in both the unweighted and the weighted cases.

Now we substitute $\rho = {1 \over k} \cdot n^{k \over {2k+2}}$. The resulting expected query time becomes $O(k  \cdot n^{{1 \over 2} + {1 \over {2k+2}}} \cdot \lambda) + O(|\hat{\Pi}|)$. We summarize the properties of our basic oracle in the following theorem.

\begin{theorem}
\label{thm:basic}
For an undirected $n$-vertex graph $G$ of arboricity $\lambda$ and a positive integer parameter $k = 1,2,\ldots$,  there exists a path-reporting distance oracle of size (whp) $O(n)$ (in addition to the size required to store the input graph $G$) that returns $(6k-1)$-approximate shortest paths $\hat{\Pi}$. The expected query time is 
$O(n^{{1 \over 2} + {1 \over {2k+2}}} \cdot k \cdot \lambda )$ in unweighted graphs and 
$O(n^{{1 \over 2} + {1 \over {2k+2}}} \cdot k \cdot( \lambda + \log n))$ in weighted ones. (The same bounds on the query time apply whp if one multiplies them by $O(\log n)$. In addition, in all cases the query time contains the additive term  $O(|\hat{\Pi}|)$.)
\end{theorem}

In particular, Theorem \ref{thm:basic} implies that for any constant $\eps > 0$ one can have a path-reporting oracle with query time $O(n^{1/2 + \eps} \lambda)$, which provides $O(1)$-approximate shortest paths for weighted undirected graphs.
Observe also that for $k=1$ we obtain a 5-approximate path-reporting oracle with query time $\tO(n^{3/4} \lambda)$.
We remark that to get the latter oracle one does not need to use the TZ oracle for the landmarks graph $\cL$. Rather one can build a DPPRO $\cH$ for all pairs of landmarks. 
(In this case $\rho = n^{1/4}$, $|L| = O(\rho)$, $|\Pairs| = |{L \choose 2}| = O(\rho^2) = O(\sqrt{n})$, and so the size of the oracle $\cH$ is $O(|\Pairs|^2 + n) = O(n)$.)

One can build the forest of SPTs rooted at the landmarks in $\tilde{O}(m)$ time. In additional $O(m \cdot \rho) = O(k \cdot m \cdot n^{1/2 - {1 \over {2k+2}}})$ time one can construct the metric closure of $L$, i.e., the graph ${\cal L}$. This graph has $n' = \rho$ vertices and $m' \le \rho^2$ edges.
In $O(k m' \cdot n'^{1/k}) = O(k \rho^{2+1/k}) = \tilde{O}(k \cdot n^{{{2k+1} \over {2k+2}}})$ time one can construct the TZ oracle for it.
To construct the DPPRO with $P = O(k \cdot \rho^{1+1/k}) = O(k \cdot n^{1/2})$ pairs one needs $O(n \cdot P^2) + \tilde{O}(k \cdot m \cdot n^{1/2 - {1 \over {2k+2}}}  ) = O(k^2 \cdot n^2)  + \tilde{O}(k \cdot m \cdot n^{1/2 - {1 \over {2k+2}}})$ time. Hence the overall construction time of this oracle is $O(k^2 \cdot n^2) + \tilde{O}(k \cdot m \cdot n^{1/2 - {1 \over {2k+2}}} )$.

In Section \ref{sec:extension} we show (see Corollary \ref{cor:extended}) that Theorem \ref{thm:basic} extends to general graphs with $m = \lambda \cdot n$ edges.

\section{An Extension to General Graphs}
\label{sec:extension}

In this section we argue that Theorem \ref{thm:basic} can be extended to general $n$-vertex graphs $G = (V,E,\omega)$ with $m = \lambda n$ edges.
In its current form the theorem only applies to graphs of arboricity at most $\lambda$. While this is sufficient for our main application, i.e., for Theorem \ref{thm:opt}, our another application (Theorem \ref{thm:EPoracle}) requires a more general result. Our extension is based on the reduction of Agarwal \etal \cite{AGHP11} of the distance oracle problem in general graphs to the same problem in bounded-degree graphs. Our argument is somewhat more general than the one from \cite{AGHP11}, as it also applies to path-reporting distance oracles. We provide our extension for the sake of completeness.

Given an $m$-edge $n$-vertex graph $G$ with $\lambda = m/n$, we split each vertex $u_i$ into $d(u) = \lceil {{\deg(u)} \over \lambda} \rceil$ copies $u^{(1)},u^{(2)},\ldots,u^{(d(u))}$. Each copy is now selected independently at random with probability $\rho/n$, for a parameter $\rho$ determined in the same way as in Section \ref{sec:basic}. The original vertex $u$ is selected to the landmarks' set if and only if at least one of its copies (which will also be called {\em virtual nodes}) is selected. Observe that the rule that we have described is equivalent to selecting $u$ with probability $d(u) \cdot {\rho \over n} = \lceil {{\deg(u)} \over \lambda } \rceil \cdot {\rho \over n}$.

The expected number of selected virtual nodes is 
$$\sum_{v \in V} d(v) \cdot {\rho \over n} ~=~ {\rho \over n} \cdot \sum_{v \in V} \lceil {{\deg(u)} \over \lambda} \rceil~ \le~ {\rho \over n}  \sum_{v \in V} ({{\deg(v)} \over \lambda} +1) ~=~ \rho + {{\rho} \over {\lambda n}}\sum_{v \in V} \deg(v) ~=~ 3\rho~.$$
The number $|L|$ of landmarks is at most the number of selected virtual nodes, and so $\Expect(|L|) \le 3\rho$. By Chernoff's bound, the number of selected virtual nodes is whp $O(\rho)$, and so, whp, $|L|^{2+2/k} = O(\rho^{2+2/k})$ as well. Hence the size of our oracle remains $O(n)$.

The rest of the construction algorithm for our distance oracle is identical to that of Section \ref{sec:basic}. (The only change is the distribution of selecting landmarks.) 
The query algorithm is identical to the query algorithm from Section \ref{sec:basic}. In particular, note that the virtual nodes have no effect on the computation, i.e., the returned paths contain only original vertices.

Next we argue that the expected query time of the modified oracle is still at most $O({n \over \rho} \cdot \lambda)$ in unweighted graphs, and $O({n \over \rho} \cdot \lambda \log n)$ in weighted ones. (As usual, we omit the additive term of the number of edges of the returned path.)
Specifically, we argue that the tests if $v \in Ball(u)$ and if $u \in Ball(v)$ can be carried out within the above expected time.

Let  $u = u_0,u_1,\ldots,u_{n-1}$ be all graph vertices ordered by a Dijkstra exploration originated from $u$, and replace each vertex $u_i$ by its $d(u_i)$ copies $u_i^{(1)},\ldots,u_i^{(d(u_i))}$. The copies appear in an arbitrary order. 
Since each virtual node has probability ${\rho \over n}$ to be selected independently of other vertices, it follows by a previous argument that the number $N$ of virtual nodes that the algorithm encounters before seeing a selected virtual node is $O({n \over \rho})$. (The algorithm actually explores only original vertices. For the sake of this argument we imagine that when the algorithm reaches a vertex $y$ it reaches its first copy $y^{(1)}$. Right after that it reaches the next copy $y^{(2)}$, etc., and then reaches $y^{(d(y))}$. After "reaching" all these copies the algorithm continues to the next original vertex.) 

Denote the original vertices explored by the algorithm $u_1,u_2,\ldots,u_{i-1},u_i$, and let $u_i^h$ be a selected copy of $u_i$. (We assume that all copies of $u_j$, for $j < i$, are not selected, and all copies $u_i^{h'}$, $h'  < h$, are also not selected.) It follows that $N = \sum_{j=1}^{i-1} d(u_j) + h$.
Hence $$\Expect\left(\sum_{j=1}^{i-1}d(u_j)\right) ~\le~ \Expect(N) ~=~ O\left({n \over \rho}\right)~.$$
 Hence $$\Expect\left( \sum_{j=1}^{i-1} \lceil {{\deg(u_j) } \over \lambda} \rceil \right) ~=~ O\left({n \over \rho}\right)$$ as well. Thus 
$$\Expect \left(\sum_{j=1}^{i-1} \deg(u_j) \right) ~=~ O \left({{\lambda n} \over \rho}\right) ~=~ O \left({m \over \rho}\right).$$

Observe that the number of edges explored by the algorithm before reaching $u_i$ is at most $\sum_{j=1}^{i-1} \deg(u_j)$. (The only  edges incident on $u_i$ 
explored by the algorithm are edges $(u_j,u_i)$, for $j < i$. These edges are accounted for in the above sum of degrees.) Hence the expected number of edges explored by the algorithm is $O({m \over \rho})$. Hence its expected running time is $O({m \over \rho})$ (respectively, $O({m \over \rho} \cdot \log n)$) in unweighted (resp., weighted) graphs. The bounds that hold with high probability are higher by a factor of $O(\log n)$.

\begin{corollary}
\label{cor:extended}
Up to constant factors, the result of Theorem \ref{thm:basic} holds for general undirected unweighted $m$-edge $n$-vertex graphs with $m = \lambda n$.
For undirected weighted graphs the expected query time becomes $O(n^{1/2 + {1 \over {2k+2}}} \cdot k \cdot \lambda \cdot \log n) = 
O(n^{1/2 + {1 \over {2k+2}}} \cdot k \cdot {m \over n} \cdot \log n)$, and the same bound applies whp if one multiplies it by another $\log n$ factor.
\end{corollary}
Since $\Expect(|L|) = O(\rho)$, the construction time of the oracle is, up to constant factors, the same as in Section \ref{sec:basic}.

This result provides a path-reporting analogue of the result of Agarwal \etal  \cite{AGHP11}, which provides stretch $O(k)$ and query time $(n\lambda)^{O(1/k)}$.
 Their oracle is not path-reporting. Our oracle is path-reporting, but its query time is significantly higher, specifically it is $n^{1/2 + O(1/k)} \cdot k \cdot \lambda$.

\section{Oracles with Smaller Query Time}
\label{sec:faster_oracles}

In this section we devise two path-reporting oracles with improved query time.
The first oracle has  size $O(m + n)$ (it stores the original graph),
and  query time $\lambda \cdot n^\eps$, for an arbitrarily small $\eps >0$. 
The stretch parameter of this oracle grows polynomially with $\eps^{-1}$. For the time being we will focus on graphs of arboricity at most $\lambda$.
The argument extends to general graphs  with $m = \lambda n$ in the same way as was described in Section \ref{sec:extension}. 
Our second oracle has size $O(n \log\log n)$ (independent of the size of the original graph) and reports stretch-$O(\log^{\stretchexp} n)$ paths
in $O(\log\log n)$ time. Both draw on techniques used in sublinear additive spanner constructions of \cite{Pet09}.
We will later build upon the first oracle to construct additional oracles that work for dense graphs as well. Like the second oracle, these later oracles will not have to store the input graph.

\subsection{Construction of an  Oracle with time $O(\lambda \cdot n^{\eps})$}
\label{sec:constr}

In this section we describe the construction algorithm of our oracle. It will use a hierarchy of landmarks' sets $L_1,L_2,\ldots,L_h$, for a positive integer parameter $h$ that will be determined later. For each index $i \in [h]$, every vertex $v$ is selected into $L_i$ independently at random with probability $p_i = {{\rho_i} \over n}$, $\rho_1 > \rho_2 > \ldots > \rho_h$. The sequence $\rho_1,\rho_2,\ldots,\rho_h$ will be determined in the sequel.
The  vertices of $L_i$ will be called the {\em $i$-level landmarks}, or shortly, the {\em $i$-landmarks}. For convenience of notation we also denote $L_0 = V$.

For each vertex $v \in V$ and index $i \in [h]$, let $\ell_i(v)$ denote the closest $i$-landmark to $v$, where ties are broken in an arbitrary consistent way.
Denote $r_i(v) = d_G(v,\ell_i(v))$ the distance between $v$ and its closest $i$-landmark $\ell_i(v)$. 
Following \cite{Pet09},  for a real number $0 < c \le 1$, 
let $\cB^c_i= \{u \mid d_G(v,u) <  c \cdot r_i(v)\}$ denote the {\em $i$th $c$-fraction-ball} of $v$.
In our analysis $c$ will be set to either 1/3 or 1. Specifically, let $\TB_i(v)$ denote the {\em one-third-ball} of $v$, 
and $\Ball_i(v) = \cB^1_i(v) = \{u \mid d_G(v,u) < r_i(v)\}$ denote the {\em $i$th ball} of $v$. 

For each vertex $v \in V$ we keep a shortest path between $v$ and $\ell_1(v)$. (This is a forest of vertex-disjoint SPTs rooted at 1-landmarks. For each 1-landmark $u'$, its SPT spans all vertices $v \in V$ which are closer to $u'$ than to any other 1-landmark.) 
Similarly, for each $i \in [h-1]$ and every $i$-landmark $u$ we keep a shortest path between $u$ and its closest $(i+1)$st landmark $\ell_{i+1}(u) = u^{(i+1)}$. Again, this entails storing a forest of vertex-disjoint SPTs rooted at $(i+1)$-landmarks, for each each index $i \in [h-1]$. Overall this part of the oracle requires $O(n \cdot h)$ space.

For the $h$th-level landmarks' set $L_h$ we build a DPPRO $\cL_h$ described in Section \ref{sec:dppro}. Given a pair $u,v$ of $h$-landmarks this oracle returns a shortest path $\Pi(u,v)$ between them in time proportional to the number of edges in this path, i.e.,  $O(|\Pi(u,v)|)$.
The space requirement of the oracle $\cL_h$ is $O(n + |L_h|^4)$, and thus we will select $\rho_h$ to ensure that $|L_h|^4 = O(n)$, i.e., $\rho_h$ will be roughly $n^{1/4}$. Denote also $\cP_h = {L_h \choose 2}$ be the set of all pairs of $h$-landmarks. 

For each index $i \in [h-1]$, we also build a DPPRO $\cD_i$ for the following set $\cP_i$ of pairs of $i$-landmarks. Each pair of $i$-landmarks $u,v$ such that either $v \in \TB_{i+1}(u)$ or $u \in \TB_{i+1}(v)$ is inserted into $\cP_i$. 

Similarly to the DPPRO $\cL_h$, given a pair $(u,v) \in \cP_i$ for some $i \in [h-1]$, the oracle $\cD_i$ returns a shortest path $\Pi(u,v)$ between $u$ and $v$ in time $O(|\Pi(u,v)|)$. 
Our oracle also stores the graph $G$ itself. We will later show a variant of this oracle that does not store $G$ (Theorem \ref{thm:fast}).
The size of the oracle is $O(n + |\Branch_i|)$, where $\Branch_i$ is the set of branching events for the set $\cP_i$. Since we aim at a linear size bound, we will ensure that $|\Branch_i| = O(n)$, for every $i \in [h-1]$. 
We will also construct a hash table $\cH_i$ for $\cP_i$ of size $O(|\cP_i|)$ that supports membership queries to $\cP_i$ in $O(1)$ time per query.
The resulting $h$-level oracle will be denoted $\Lambda_h$.

\comment
\inline Remark: There will be later a few variants of this oracle that also contain an oracle $\cD_0$ for pairs $\cP_0$ of 0-landmarks (i.e., $L_0 = V$) $u,v$ which satisfy the same rule (i.e., $v \in \TB(u)$ or $u \in \TB(v)$). However, the current version of the oracle does not contain $\cD_0$. 
\commentend

\subsection{The Query Algorithm}
\label{sec:query}

Next, we describe the query algorithm of our oracle $\Lambda_h$. The query algorithm is given a pair $u = u^{(0)}, v = v^{(0)}$ of vertices. The algorithm starts with testing if $u \in \Ball_1(v)$ and if $v \in \Ball_1(u)$. For this test the algorithm just conducts a Dijkstra search from $v$ until it discovers  either $v^{(1)}$ or $u$ (and, symmetrically, also conducts a search from $u$).

Observe that by Equation (\ref{eq:geom_expect}), the expected size of $\Ball_1(v)$ and of $\Ball_1(u)$ is $O({n \over {\rho_1}})$, and whp both these sets have size 
$O({n \over {\rho_1}} \cdot \log n)$. Hence the running time of this step is, whp, $\tO({n \over {\rho_1}} \cdot \lambda)$.  (Specifically, it is $O({n \over {\rho_1}} \cdot \lambda \cdot \log n)$ in unweighted graphs, and $O({n \over {\rho_1}} \cdot \log n \cdot (\lambda + \log n))$ in weighted ones.  The expected running time of this step is smaller by a factor of $\log n$ than the above bound.)

If the algorithm discovers that $v \in \Ball_1(u)$ or that $u \in \Ball_1(v)$ then it has found the shortest path between $u$ and $v$. In this case the algorithm returns this path. Otherwise it has found $u^{(1)} = \ell_1(u^{(0)})$ and $v^{(1)} = \ell_1(v^{(0)})$. 

In general consider a situation when for some index $j$, $1 \le j \le h$, the algorithm has already computed $u^{(j)}$ and $v^{(j)}$. In this case, inductively, the algorithm has already computed shortest paths $\Pi(u^{(0)},u^{(1)}), \Pi(u^{(1)},u^{(2)}),\ldots,\Pi(u^{(j-1)},u^{(j)})$ and 
 $\Pi(v^{(0)},v^{(1)}), \Pi(v^{(1)},v^{(2)}),\ldots,\Pi(v^{(j-1)},v^{(j)})$  between $u^{(0)}$ and $u^{(1)}$, $u^{(1)}$ and $u^{(2)}$, $\ldots$, $u^{(j-1)}$ and $u^{(j)}$, $v^{(0)}$ and $v^{(1)}$, $v^{(1)}$ and $v^{(2)}$, $\ldots$, $v^{(j-1)}$ and $v^{(j)}$, respectively. (Note that the base case $j=1$ has been just argued.)

For $j < h$, the query algorithm of our oracle $\Lambda_h$ then queries the hash table $\cH_j$ whether the pair $(u^{(j)},v^{(j)}) \in \cP_j$. If it is the case then the algorithm queries the oracle $\cD_j$, which, in turn, returns the shortest path $\Pi(u^{(j)},v^{(j)})$ between $u^{(j)}$ and $v^{(j)}$ in time $O(|\Pi(u^{(j)},v^{(j)})|)$.
The algorithm then reports the concatenated path 
\begin{eqnarray*}
\Pi(u,v) &=& \Pi(u^{(0)},u^{(1)}) \cdot   \Pi(u^{(1)},u^{(2)}) \cdot \ldots  \Pi(u^{(j-1)},u^{(j)}) \cdot \Pi(u^{(j)},v^{(j)}) \\
&\cdot & \Pi(v^{(j)},v^{(j-1)}) \cdot 
 \ldots   \cdot \Pi(v^{(2)},v^{(1)}) \cdot \Pi(v^{(1)},v^{(0)})~.
\end{eqnarray*}
Computing this concatenation requires $O(j) \le O(|\Pi(u,v)|)$ time.

In the complementary case when $(u^{(j)},v^{(j)}) \nin \cP_j$, the algorithm fetches the prerecorded paths $\Pi(u^{(j)},u^{(j+1)})$ and  $\Pi(v^{(j)},v^{(j+1)})$, and invokes itself recursively on the pair $(u^{(j+1)},v^{(j+1)})$.  (Recall that for each index $j$, $1 \le j \le h-1$, the algorithm stores a forest of vertex-disjoint SPTs rooted at $(j+1)$-landmarks $L_{j+1}$. These SPTs enable us to compute the paths $\Pi(u^{(j)},u^{(j+1)})$, $\Pi(v^{(j)},v^{(j+1)})$ for all $j \in [h-1]$, in time proportional to the number of edges in these paths.) 

Finally, if $j = h$ then we query the DPPRO $\cL_h$ of the graph $L_h$ with the query $(u^{(h)},v^{(h)})$. 
(Note that it is not necessary to query if $(u^{(h)},v^{(h)})$ is in the DPPRO $\cL_h$, since, by construction, all such pairs are there.)
The query returns the shortest path 
 between them in time $O(|\Pi(u^{(h)},v^{(h)})|)$. It follows that the overall running time of the query algorithm is dominated by the time required to compute $\Pi(u^{(0)},u^{(1)})$ and $\Pi(v^{(0)},v^{(1)})$. Specifically, it is
$$\tO({n \over {\rho_1}} \cdot \lambda) + \sum_{i=0}^{j-1} \left(|\Pi(u^{(i)},u^{(i+1)})| + |\Pi(v^{(i)},v^{(i+1)})|\right) + |\Pi(u^{(j)},v^{(j)})|,$$
 where $1 \le j \le h$ is the smallest index such that $(u^{(j)},v^{(j)}) \in \cP_j$.
(Recall that for $j=h$, $\cP_h = {L_h \choose 2}$, i.e., all pairs of $h$-landmarks belong to $\cP_h$.)
Hence the overall query time is $\tO({n \over {\rho_1}} \cdot \lambda)  + O(|\Pi(u,v)| + h)$, where $\Pi(u,v)$ is the path that the algorithm ultimately returns. 

\inline Remark: If for each index $0 \le j \le h-1$ at least one of the subpaths $\Pi(u^{(j)},u^{(j+1)}), \Pi(v^{(j)},v^{(j+1)})$ is not empty then $h \le |\Pi(u,v)|$, and the resulting query time is $\tO({n \over {\rho_1}} \lambda) + O(|\Pi(u,v)|)$. One can artificially guarantee that all these subpaths will not be empty, i.e., that $u^{(j)} \neq u^{(j+1)}$ and $v^{(j)} \neq 
v^{(j+1)}$,
for every $j$. To do this one can modify the construction slightly so that  the set of $i$-landmarks and the set of $j$-landmarks will be  disjoint for all $i \ne j$.
Under this modification of the algorithm the query time is $\tO({n \over {\rho_1}} \cdot \lambda) + O(|\Pi(u,v)|)$, while the stretch guarantee of the oracle (which will be analyzed in Section \ref{sec:stretch}) stays the same. This modification can make oracle's performance only worse than it is without this modification, but the bounds on the query time of the modified oracle in terms of the number of edges in the returned path become somewhat nicer. (See Theorem \ref{thm:fast}.)

\subsection{The Stretch Analysis}
\label{sec:stretch}

Recall that in the case that $v \in \Ball_1(u)$ or $ u \in \Ball_1(v)$ our algorithm returns the exact shortest path between $u=u^{(0)}$ and $v = v^{(0)}$. Hence we next consider the situation when $v \nin \Ball_1(u)$ and $u\nin \Ball_1(v)$. 
For brevity let $d = d^{(0)} = d_G(u,v)$.
At this point the algorithm also has already computed $u^{(1)}$ and $v^{(1)}$, along with the shortest paths $\Pi(u^{(0)},u^{(1)})$ and $\Pi(v^{(0)},v^{(1)})$ between $u^{(0)}$ and $u^{(1)}$ and between $v^{(0)}$ and $v^{(1)}$, respectively.
Observe that in this scenario we have $d_G(u^{(0)},u^{(1)}),d_G(v^{(0)},v^{(1)}) \le d$, and so 
$$d_G(u^{(1)},v^{(1)}) \le d_G(u^{(1)},u^{(0)}) + d_G(\uzero,\vzero) +d_G(\vzero,\vone) \le 3 \cdot d.$$
Hence if $(\uone,\vone) \in \cP_1$ then the path $\Pi(\uzero,\uone) \cdot \Pi(\uone,\vone) \cdot \Pi(\vone,\vzero)$  returned by the algorithm is a 5-approximate path between $u$ and $v$. Indeed, its length is at most
$$d_G(\uzero,\uone) + d_G(\uone,\vone) + d_G(\vone,\vzero) ~\le~ d+ 3\cdot d + d ~= ~ 5\cdot d.$$
More generally, suppose the query algorithm reached the $j$-level landmarks $\uj,\vj$, for some $j$, $1 \le j \le h-1$, and suppose that $(\uj,\vj) \nin \cP_j$.
This means that $\vj \nin \TB_{j+1}(\uj)$ and $\uj \nin \TB_{j+1}(\vj)$.
By definition of the one-third-ball it follows that 
$$d_G(\uj,\vj) ~\ge~ {1 \over 3} \cdot d_G(\uj,u^{(j+1)}) 
~= ~ {1 \over 3} \cdot r_{j+1}(u^{(j)})~,$$
and
$$ d_G(\uj,\vj)  ~\ge~ {1 \over 3} \cdot d_G(\vj,v^{(j+1)}) ~= ~ {1 \over 3}\cdot r_{j+1}(v^{(j)})~,$$
where $u^{(j+1)}$ (respectively, $v^{(j+1)}$) is the $(j+1)$-landmark closest to $\uj$ (resp., $\vj$).

Hence 
$$d_G(u^{(j+1)},v^{(j+1)}) ~\le~ \dG(u^{(j+1)},\uj) + \dG(\uj,\vj) + \dG(\vj,v^{(j+1)}) ~\le~ 7 \cdot \dG(\uj,\vj)~.$$
Denote by $p$, $1 \le p \le h$, the index for which the algorithm discovers that $(u^{(p)},v^{(p)}) \in \cP_p$. 
(Since $(u^{(h)},v^{(h)}) \in \cP_h$ for every pair $(u^{(h)},v^{(h)})$ of $h$-landmarks, it follows that 
the index $p$ is well-defined.)

 We have seen that $\dG(\uone,\vone) \le 3d$, and for every index $j$, $1 \le j \le p -1$,
$\dG(u^{(j+1)},v^{(j+1)}) \le 7 \cdot \dG(\uj,\vj)$. Hence for every $j$, $ 1 \le j \le p$, it holds that
$\dG(\uj,\vj) \le 3 \cdot 7^{j-1} \cdot d$.
Denote $\dj = 3 \cdot 7^{j-1} \cdot d$, for $0 \le j \le p$.
Also, $\dG(\uzero,\uone),d_G(\vzero,\vone) \le d = \dzero$, and for every index $j$, $1 \le j \le p-1$, 
$$\dG(\uj,u^{(j+1)}) ~\le~ 3 \cdot \dG(\uj,\vj) ~\le 3 \cdot \dj ~= ~ 3^2 \cdot 7^{j-1} \cdot d~.$$
Hence the length of the path 
$$\Pi(\uzero,\uone) \cdot \ldots \cdot \Pi(u^{(p-1)},u^{(p)}) \cdot \Pi(u^{(p)},v^{(p)}) \cdot \Pi(v^{(p)},v^{(p-1)}) \cdot \ldots \Pi(\vone,\vzero)$$ returned by the algorithm is at most
\begin{eqnarray*}
&&\dzero + 3 \cdot \left(\sum_{j=1}^{p-1} \dj \right) +d^{(p)} + 3 \cdot \left( \sum_{j=1}^{p-1} \dj \right) + \dzero ~ = ~ \\
&& d\cdot \left(2 \cdot \left(1 + 3 \cdot \left( \sum_{j=1}^{p-1} 3 \cdot 7^{j-1} \right)\right) + 3\cdot 7^{p-1}\right) ~=~ d\cdot (6 \cdot 7^{p-1} - 1)~.
\end{eqnarray*}
Since $p \le h$ we conclude that the oracle has stretch at most $6 \cdot 7^{h-1} - 1$.

\subsection{The Size of the Oracle}
\label{sec:size}

For each index $i \in [h]$, our oracle stores a forest of (vertex-disjoint) SPTs rooted at $i$-landmarks. Each of these forests requires $O(n)$ space, i.e., together these $h$ forests require $O(n \cdot h)$ space.

We next set the values $\rho_1 > \rho_2 > \ldots > \rho_h$ so that each of the auxiliary oracles $\cD_1,\cD_2,\ldots,\cD_{h-1},\cL_h$ requires $O(n)$ space.
Each of the hash tables $\cH_1,\cH_2,\ldots,\cH_h$ associated with these oracles requires less space than its respective oracle. Recall that the parameter $\rho_1$ also determines the query time. (It is $\tO({n \over {\rho_1}} \lambda) + O(|\Pi|)$, where $\Pi$ the path returned by the algorithm. In the sequel we will often skip the additive term of $O(|\Pi|)$ when stating the query time.)

For each $i \in [h]$ we write $\rho_i = n^{\alpha_i}$, where $\alpha_i = 1 - (3/4)^{h-i+1}$.
Observe that $\alpha_h = 1/4$, i.e., $\rho_h = n^{1/4}$.

Hence $\Expect(|L_h|) = \rho_h =  n^{1/4}$, and by Chernoff's bound, whp, $|L_h| = O( n^{1/4})$. (Recall that $|L_h|$ is a Binomial random variable.)
Hence the DPPRO $\cL_h$ for $\cP_h = {L_h \choose 2}$ requires space $O(|L_h|^4 + n) = O(n)$, whp.

Next we analyze the space requirements of the oracles $\cD_1,\cD_2,\ldots,\cD_{h-1}$. Fix an index $i \in [h-1]$, and recall that the space requirement of the DPPRO $\cD_i$ is $O(n + |\Branch_i| + |\cP_i|)$, where $\Branch_i$ is the set of branching events for the set $\cP_i$ of pairs of vertices. Next we argue that (whp) $|\Branch_i| = O(n)$.
Recall that the set $\cP_i$ contains all pairs of $i$-landmarks $(u^{(i)},v^{(i)})$ such that either $v^{(i)} \in \TB_{i+1}(u^{(i)})$ or 
 $u^{(i)} \in \TB_{i+1}(v^{(i)})$. 

The following two lemmas from \cite{Pet09} are the key to the analysis of the oracle's size. The first says that with our definition of $\cP_{i+1}$ all branching events are confined to $(i+1)$st level balls. The second bounds the expected number of branching events in terms of the sampling probabilities.
For completeness, the proofs of these lemmas are provided in Appendix \ref{sec:missing_proofs}.

\begin{lemma}
\label{lm:quadruples}
Suppose that $v \in \TB_{i+1}(u)$. Then if $(x,y) \in \cP_{i+1}$ and there is a branching event between the pairs $(u,v)$ and $(x,y)$ then necessarily $x,y \in \Ball_{i+1}(u)$.
\end{lemma}
\def\PFquadlemma
{
Suppose for contradiction that there exists a pair $(x,y) \in \cP_{i+1}$ such that the pairs $(u,v),(x,y)$  participate in a branching event $\beta$, and such that either $x \nin \Ball_{i+1}(u)$ or $y \nin \Ball_{i+1}(u)$. Then $\beta = (\Pi(u,v),\Pi(x,y),z)$, where $\Pi(u,v)$ (respectively, $\Pi(x,y)$) is a shortest path between $u$ and $v$ (respectively, between $x$ and $y$), and $z$ is a node at which these two paths branch.
Since $(x,y) \in \cP_{i+1}$ it follows that either $y \in \TB_{i+1}(x)$ or $x \in \TB_{i+1}(y)$. Without loss of generality suppose that $ y \in \TB_{i+1}(x)$.

The proof splits into two cases. In the first case we assume that $x \nin \Ball_{i+1}(u)$, and in the second we assume that $y \nin \Ball_{i+1}(u)$. 
(Note that roles of $x$ and $y$ are not symmetric.)
In both cases we reach a contradiction.

We start with the case $x \nin \Ball_{i+1}(u)$.
Observe that
$\dG(x,z) \le \dG(x,y) < \third \cdot r_{i+1}(x)$ and  $d_G(u,z) \le d_G(u,v) < \third \cdot r_{i+1}(u)$.
Denote $\delta = \dG(u,u^{(i+1)}) = r_{i+1}(u)$, where $u^{(i+1)} = \ell_{i+1}(u)$.
Denote also $\delta' = \dG(u,x)$.
Observe that $r_{i+1}(x) \le \dG(x,u^{(i+1)}) \le \delta + \delta'$, and also (since $x \nin \Ball_{i+1}(u)$) $\delta' = d_G(u,x) \ge \delta = r_{i+1}(u)$.
Then
$$
\dG(u,z) + \dG(z,x) ~< ~ \third \cdot r_{i+1}(u) + \third \cdot r_{i+1}(x) ~\le~ {\delta \over 3} + \third \cdot (\delta  + \delta') 
~ \le ~ \delta' ~=~ \dG(u,x)~.
$$
Hence $\dG(u,z) + \dG(z,x) < \dG(u,x)$, contradicting the triangle inequality.

We are now left with the case that $x \in \Ball_{i+1}(u)$, but $ y \nin \Ball_{i+1}(u)$. Then 
$\dG(y,z) \le \dG(x,y) <  \third \cdot r_{i+1}(x)$. Also, $\dG(u,z) \le \dG(u,v) < \third \cdot r_{i+1}(u)$. 
In addition, $r_{i+1}(x) \le d_G(x,u^{(i+1)}) \le d_G(x,u) + r_{i+1}(u) \le 2\delta$.
(Note that $d_G(x,u) \le \delta = r_{i+1}(u)$, because $x \in \Ball_{i+1}(u)$.) 
Hence
$$\dG(u,z) + \dG(z,y) ~<~ \third \cdot (r_{i+1}(u) + r_{i+1}(x)) ~\le~ \third \cdot (\delta + 2\delta) ~=~ \delta ~\le~ \dG(u,y)~.$$
(The last inequality is because, by an assumption, $y \nin \Ball_{i+1}(u)$.)
This is, however, again a contradiction to the triangle inequality.
\QED
}


\begin{lemma}
\label{lm:branch}
Whp, $|\Branch_i| = O\left({{\rho_i^4} \over {\rho_{i+1}^3}} \cdot \log^3 n\right)$, and $\Expect(|\Branch_i|) = O\left({{\rho_i^4} \over {\rho_{i+1}^3}}\right)$. Moreover, whp $|\cP_i| =  O\left({{\rho_i^2} \over {\rho_{i+1}}} \cdot \log n\right)$, and 
$\Expect(|\cP_i|) =  O\left({{\rho_i^2} \over {\rho_{i+1}}} \right)$.
\end{lemma}
\def\PFbranchlemma
{
Recall that (see \cite{CE05}, Lemma 7.5) each pair $(u,v), (x,y)$ may produce at most two branching events. Hence next we focus on providing an upper bound on the number of intersecting pairs of paths $\Pi(u,v), \Pi(x,y)$ for $(u,v),(x,y) \in \cP_i$.

By the previous lemma, for a pair $(u,v),(x,y)$ to create a branching event  there must be one of these four vertices (without loss of generality we call it $u$) such that the three other vertices belong to $\Ball_{i+1}(u)$. Hence the number of intersecting pairs as above is at most (a constant factor multiplied by) the number of quadruples $(u,v,x,y)$ with $v,x,y \in \Ball_{i+1}(u)$. For a fixed $i$-landmark $u$, the number of vertices in its $(i+1)$st ball $\Ball_{i+1}(u^{(i)})$ is, whp, $O\left({n \over {\rho_{i+1}}} \cdot \log n\right)$. (This random variable is distributed geometrically with the parameter $p = {{\rho_{i+1}} \over n}$.)
Each of the vertices in $\Ball_{i+1}(u)$ has probability ${{\rho_i} \over n}$ to belong to $L_i$, independently of other vertices.
Hence, by Chernoff's bound, whp, there are ${{\rho_i} \over n} \cdot O\left({n \over {\rho_{i+1}}} \cdot \log n\right) = O\left({{\rho_i} \over {\rho_{i+1}}} \cdot \log n\right)$ $i$-landmarks in $\Ball_{i+1}(u)$. (We select the constant $c$ hidden by the $O$-notation in  $O\left({n \over {\rho_{i+1}}} \cdot \log n\right) $ to be sufficiently large. Then the expectation is $c \cdot {{\rho_i} \over {\rho_{i+1}}} \cdot \log n \ge c \cdot \log n$. Hence the Chernoff's bound applies with high probability.) 

Hence the number of triples $v,x,y$ of $i$-landmarks in $\Ball_{i+1}(u)$ is, whp, $O\left({{\rho_i^3} \over {\rho_{i+1}^3}} \cdot \log^3 n\right)$.
The number of $i$-landmarks $u$ is, by the Chernoff's bound, whp, $O(\rho_i)$.
Hence the number of quadruples as above is, whp, at most 
$$O(\rho_i) \cdot O\left({{\rho_i^3} \over {\rho_{i+1}^3}} \cdot \log^3 n\right) = O\left({{\rho_i^4} \over {\rho_{i+1}^3}} \cdot \log^3 n\right)~.$$
Also, the number of pairs $|\cP_i|$ is at most the number of $i$-landmarks (whp, it is $O(\rho_i)$) multiplied by the maximum number of $i$-landmarks in an $(i+1)$-level ball $\Ball_{i+1}(u)$ (whp, it is  $O\left({{\rho_i} \over {\rho_{i+1}}} \cdot \log n\right)$), i.e., $|\cP_i| =  O\left({{\rho_i^2} \over {\rho_{i+1}}} \cdot \log n\right)$.

Next we argue that the expected number of quadruples $(u,v,x,y)$ of $i$-landmarks such that $v,x,y \in \Ball_{i+1}(u)$ is $O\left({{\rho_i^4} \over {\rho_{i+1}^3}} \right)$ and that $\Expect(|\cP_i|) = O\left({{\rho_i^2} \over {\rho_{i+1}}}\right)$.

For a fixed vertex $u$, write $X(u) = I(\{ u \in L_i\}) \cdot Y(u)$, where $Y(u)$ is the number of triples of distinct $i$-landmarks different from $u$  which belong to $\Ball_{i+1}(u)$, and $I(\{u \in L_i\})$ is the indicator random variable of the event $\{u \in L_i\}$. (Note that the ball is defined even if $u \nin L_i$.)
Observe that the random variables $I(\{u \in L_i\})$ and $Y(u)$ are independent, and thus
$$\Expect(X(u)) ~=~ \Expect(I(\{u \in L_i\})) \cdot \Expect(Y(u)) ~=~ {{\rho_i} \over n} \cdot \Expect(Y(u))~.$$
Let $\sigma = (v_1,v_2,\ldots,v_{n-1})$ be the sequence of vertices ordered by the non-decreasing distance from $u$. (They appear in the  order in which the Dijkstra algorithm initiated at $u$ discovers them.)
For $k = 3,4,\ldots,n-1$, denote by $\cJ_k$ the random variable which is equal to 0 if $v_{k+1}$ is not the first vertex in $\sigma$ which belongs to $L_{i+1}$. If $v_{k+1}$ is the first vertex as above then $\cJ_k$  is equal to the number of triples $v_{j_1},v_{j_2},v_{j_3}$, $1 \le j_1 < j_2 < j_3 \le k$ such that $v_{j_1},v_{j_2},v_{j_3} \in L_i$. 
Also, for each quadruple $1 \le j_1 < j_2 < j_3 < j_4 \le n-1$ of indices,  define $J(j_1,j_2,j_3,j_4)$ to be the indicator random variable of the event that $v_{j_1},v_{j_2},v_{j_3} \in L_i$,  $v_{j_4} \in L_{i+1}$, and for each $j$, $1 \le j < j_4$, the vertex $v_j$ is not an $(i+1)$-landmark. Observe that 
$$\Expect(J(j_1,j_2,j_3,j_4)) = \left({{\rho_i} \over n}\right)^3 \cdot \left(1 - {{\rho_{i+1}} \over n}\right)^{j_4-1} \cdot {{\rho_{i+1}} \over n}~.$$
Also, 
$$\Expect(\cJ_k) ~=~ \sum_{1 \le j_1 < j_2 < j_3 \le k} \Expect(J(j_1,j_2,j_3,k+1)) ~=~ {k \choose 3} \left({{\rho_i} \over n}\right)^3 \cdot \left(1 - {{\rho_{i+1}} \over n}\right)^k \cdot {{\rho_{i+1}} \over n}~.$$ 
Note that $Y(u) = \sum_{k=3}^{n-2} \cJ_k$, and so
$$\Expect(Y(u)) ~\le~ \sum_{k=3}^\infty {k \choose 3} \left({{\rho_i} \over n}\right)^3 \cdot \left(1 - {{\rho_{i+1}} \over n}\right)^k \cdot {{\rho_{i+1}} \over n}~.$$
Denote $A = 10 {n \over {\rho_{i+1}}}$. For $k \le A$, since $(1 - {{\rho_{i+1}} \over n})^k = O(1)$, it follows that
$$\sum_{k=3}^A {k \choose 3} \left({{\rho_i} \over n}\right)^3 \cdot \left(1 - {{\rho_{i+1}} \over n}\right)^k \cdot {{\rho_{i+1}} \over n}
~=~ O\left({{\rho_i^3 \cdot \rho_{i+1}} \over n^4} \right) \sum_{k=3}^{A} k^3 ~=~ O\left({{\rho_i^3 } \over {\rho_{i+1}^3}}\right)~.$$
 Also,
$$
\sum_{k=A+1}^\infty {k \choose 3} \left({{\rho_i} \over n}\right)^3 \cdot \left(1 - {{\rho_{i+1}} \over n}\right)^k \cdot {{\rho_{i+1}} \over n} ~ \le ~
O\left({{\rho_i^3 \cdot \rho_{i+1}} \over {n^4}} \right) \cdot \sum_{k=A+1}^\infty k^3 \cdot \left(1 - {{\rho_{i+1}} \over n}\right)^k~.$$
Denote $\gamma = 1 - \rho_{i+1}/n$.
Then
$$ \sum_{k=A+1}^\infty k^3 \gamma^k ~\le~ \frac{\mathrm{d}^3}{\mathrm{d} \gamma^3} \sum_{k=A+1}^\infty \gamma^{k+3} ~\le ~ {{\mathrm{d}^3} \over {\mathrm{d}\gamma^3}} {1 \over {1- \gamma}} ~=~ {6 \over {(1 - \gamma)^4}} ~= ~ O\left(\left({n \over {\rho_{i+1}}}\right)^4\right)~.$$

Hence 
$$
\sum_{k=A+1}^\infty {k \choose 3} \left({{\rho_i} \over n}\right)^3 \cdot \left(1 - {{\rho_{i+1}} \over n}\right)^k \cdot {{\rho_{i+1}} \over n} ~ = ~ 
O\left({{\rho_i^3 \cdot \rho_{i+1}} \over {n^4}} \right) \cdot O\left( \left({n \over {\rho_{i+1}}}\right)^4\right)~ =~ O\left({{\rho_i^3 } \over {\rho_{i+1}^3}}\right)~,$$
and so $\Expect(Y(u)) = O({{\rho_i^3} \over {\rho_{i+1}^3}})$. 
Hence $\Expect(X(u)) = {{\rho_i} \over n} \cdot \Expect(Y(u)) = O({{\rho_i^4} \over {\rho_{i+1}^3}} \cdot {1 \over n})$.

Finally, the overall expected number of quadruples $(u,v,x,y)$ of $i$-landmarks such that $v,x,y \in \Ball_{i+1}(u)$ is, by linearity of expectation, at most $\sum_{v \in V} \Expect(X(u)) = O({{\rho_i^4} \over {\rho_{i+1}^3}})$.

A similar argument provides an upper bound of $O\left({{\rho_i^2} \over {\rho_{i+1}}}\right)$ on the expected number of pairs $|\cP_i|$.
We shortly sketch it below.

For a vertex $u$, let $X'(u) = I(\{u \in L_i\}) \cdot Y'(u)$, where $Y'(u)$ is the number of $i$-landmarks which belong to $\Ball_{i+1}(u)$.
Clearly, $\Expect( I(\{u \in L_i\}))  = \rho_i/n$, and the two random variables ($ I(\{u \in L_i\})$ and $Y'(u)$) are independent.
For every integer $k \ge 1$, let $\cJ'_k$ be a random variable which is equal to 0 if $v_{k+1}$ is not the first vertex in $\sigma$ which belongs to $L_{i+1}$. Otherwise it is the number of $i$-landmarks among $v_1,v_2,\ldots,v_k$.
For integer $j_1,j_2$, $1 \le j_1 < j_2 \le n-1$, let $J'(j_1,j_2)$ be the indicator random variable of the event that $v_{j_1} \in L_i$, $v_{j_2} \in L_{i+1}$, and for every $j < j_2$, it holds that $v_j \nin L_{i+1}$.  Then 
$$\Expect(J'(j_1,j_2)) ~=~ {{\rho_i} \over n} \cdot \left(1 - {{\rho_{i+1}} \over n}\right)^{j_2-1} \cdot {{\rho_{i+1}} \over n}~.$$
Hence
$$\Expect(\cJ'_k) ~=~ \sum_{1 \le j_1 \le k} \Expect(J'(j_1,k+1)) = {{\rho_i \cdot \rho_{i+1}} \over {n^2}} \cdot k \cdot \left(1 - {{\rho_{i+1}} \over n}\right)^k~,$$
and
$$\Expect(Y'(u)) ~\le~ \sum_{k=1}^\infty \Expect(\cJ'_k) ~=~ {{\rho_i \cdot \rho_{i+1}} \over {n^2}} \cdot \sum_{k=1}^\infty  k \cdot \left(1 - {{\rho_{i+1}} \over n}\right)^k~.$$
Write $A = 10 {n \over {\rho_{i+1}}}$, and 
$$\sum_{k=1}^\infty k \left(1 - {{\rho_{i+1 }} \over n}\right)^k = \sum_{k=1}^A  k \left(1 - {{\rho_{i+1}} \over n}\right)^k  + \sum_{k > A}  k \left(1 - 
{{\rho_{i+1}}  \over n} \right)^k~.$$ 
Each term of the first sum is $O(1)$, and thus the first sum is at most $O(A^2) = O(n^2/\rho_{i+1}^2)$.
The second sum is at most ${d \over {d\gamma}} \sum_{k > A} \gamma^{k+1} \le {d \over {d\gamma}} {1 \over {1 - \gamma}} = O(n^2/\rho_{i+1}^2)$ as well.
Hence $$\Expect(Y'(u)) ~=~ {{\rho_i \cdot \rho_{i+1}} \over {n^2}} \cdot  O\left({{n^2} \over {\rho_{i+1}^2}}\right) = O\left({{\rho_i} \over {\rho_{i+1}}}\right)~.$$
Hence $\Expect(X'(u)) = O(\rho_i^2/(\rho_{i+1} n))$, and
by linearity of expectation we conclude that $\Expect(|\cP_i|) \le \sum_{u \in V} \Expect(X'(u)) =  O(\rho_i^2/\rho_{i+1} )$.
\QED
}

Observe that with our choice of $\rho_i$ ($\rho_i = n^{\alpha_i}$, $\alpha_i = 1 - (3/4)^{h-i+1}$, 
 for every $i \in [h]$), it holds for every $i \in [h-1]$ that
$O\left({{\rho_i^4 } \over {\rho_{i+1}^3}} \right) ~=~ O(n^{4\alpha_i - 3\alpha_{i+1}} ) ~=~ O(n)$, and $O\left({{\rho_i^2} \over {\rho_{i+1}}}\right) = O(n^{2\alpha_i - \alpha_{i+1}}) = O(n^{1 - {1 \over 2} ({3 \over 4})^{h-i}})$.
Hence by Lemma \ref{lm:branch}, for each $i \in [h-1]$, the oracle $\cD_i$ requires expected space $O(n+|\Branch_i| + |\cP_i|) = O(n)$.
Thus the overall expected space required by our $h$-level oracle oracle $\Lambda_h$ (in addition to the space required to store the original graph $G$) is $O(n \cdot h)$.
Recall that the query time is (whp) $\tO((n/\rho_1) \lambda ) = \tO(n^{(3/4)^h} \cdot \lambda)$.

The argument described in Section \ref{sec:extension} enables us to extend these results to general $m$-edge $n$-vertex graphs. 

\begin{theorem}
\label{thm:multilevel}
For any parameter $h = 1,2,\ldots$ and any $n$-vertex undirected possibly weighted graph $G$ with arboricity $\lambda$, the path-reporting distance oracle $\Lambda_h$  uses expected space $O(n \cdot h)$, in addition to the space required to store $G$.
Its stretch is $(6 \cdot 7^{h-1} - 1)$, and its query time is (whp) $\tO(n^{(3/4)^h} \lambda)$.
The same result applies for any $m$-edge $n$-vertex graph with $\lambda = m/n$.    
\end{theorem}

Specifically, in unweighted graphs with arboricity $\lambda$ the query time is $O((n/\rho_1) \cdot \lambda \cdot  \log n) = O(n^{(3/4)^h} 
\cdot \lambda \cdot  \log n)$, while in weighted graphs 
it is $ O(n^{(3/4)^h} \cdot (\lambda + \log n) \log n)$.
In unweighted $m$-edge $n$-vertex graphs the query time is $O(n^{(3/4)^h} \cdot {m \over n} \cdot \log n)$,  while in $m$-edge $n$-vertex weighted graphs it is $O(n^{(3/4)^h} \cdot {m \over n} \cdot \log^2 n)$.

By introducing a parameter $t = (4/3)^h$ we get query time $\tO(n^{1/t} \lambda)$, space $O(n \cdot \log t)$, and stretch at most 
$t^{\log_{4/3} 7}$. (The  exponent is $\approx 6.76$.)

\begin{corollary}
\label{cor:multilevel}
For any constant $t$ of the form $t = (4/3)^h$ (for a positive integer $h$) and an $n$-vertex graph $G$ with arboricity $\lambda$, our path-reporting distance oracle $\Lambda_h$ uses expected space $O(n)$ (in addition to the space needed to store $G$). It provides stretch at most $t^{\log_{4/3} 7}$, and its query time is (whp) $\tO(n^{1/t} \lambda)$.
(For a non-constant $t$ the space requirement becomes $O(n \cdot \log t)$.)
The same result applies for any $m$-edge $n$-vertex graph with $\lambda = m/n$.
\end{corollary}

Yet better bounds can be obtained if one is interested in small {\em expected}  query time.
The expected query time is dominated by the time required to test if $v \in \Ball_1(u)$ and if $u \in \Ball_1(v)$. For unweighted graphs these tests require $O({n \over {\rho_1}} \lambda) = O(n^{(3/4)^h} \lambda)$ expected time. 

\begin{corollary}
\label{cor:expected}
For any  $t$ of the form $t = (4/3)^h$, for a positive integer $h$, and an $n$-vertex $m$-edge graph $G$, our path-reporting oracle $\Lambda_h$ uses expected  $O(n \cdot h)$ space  in addition to the space required to store $G$. It provides stretch at most $t^\stretchexp$, and its expected query time is $O(n^{1/t} \cdot (m/n) + \log t)$ for unweighted graphs. In the case of weighted graphs the expected query time is $O(n^{1/t} (m/n) \cdot \log n)$.
\end{corollary}

Consider now the oracle $\Lambda_h$ for a superconstant number of levels $h = \log_{4/3}(\log n + 1)$.
Then $\rho_1 = (2n)^{\alpha_1} = n$. In other words, all vertices $V$ of $G$ are now defined as the first level landmarks (1-landmarks), i.e., $L_1 = V$.
(For levels $i=2,3,\ldots,h$, landmarks $L_i$ are still selected at random from $V$ with probability $\rho_i/n < 1$, independently. For level 1 this probability is 1.)
Recall that our oracle starts with testing if $v \in \Ball_1(u)$ and if $u \in \Ball_1(v)$. Now both these balls are empty sets, because all vertices belong to $L_1$.
Thus with this setting of parameters the oracle $\Lambda_h$ no longer needs to conduct this time-consuming test. Rather it proceeds directly to querying the oracle $\cD_1$. Remarkably, this variant of our oracle does not require storing the graph $G$. (Recall that the graph was only used by the query algorithm for testing  
 if $v \in \Ball_1(u)$ and if $u \in \Ball_1(v)$.)
The query time of the new oracle is now dominated by the $h$ queries to the oracles $\cD_1,\cD_2,\ldots,\cD_{h-1},\cL_h$, i.e., $O(h) = O(\log\log n)$.
Recall that, by the remark at the end of Section \ref{sec:query}, one can always make our oracle to return paths with at least $h$ edges, and thus the $O(h) = O(\log\log n)$ additive term in the query time can be swallowed by $O(|\Pi|)$, where $\Pi$ is the path that our oracle returns.

Denote by $\tLambda$ the oracle which was just described. The  stretch of $\tLambda$ is (by Theorem \ref{thm:multilevel})  $6 \cdot 7^{h-1} - 1 = O(\log^\stretchexp n)$. 

\begin{theorem}
\label{thm:fast}
The oracle $\tLambda$ is a path-reporting oracle with expected space $O(n \log\log n)$, where $n$ is the number of vertices of its input undirected  weighted graph $G$. Its stretch is $O(\log^\stretchexp n)$
and its query time is $O(\log\log n)$. (It can be made $O(1)$, but the paths returned by the oracle will then contain $\Omega(\log\log n)$ edges.)
\end{theorem}

Note that by Markov's inequality, Theorem \ref{thm:fast} implies that one can produce a path-reporting oracle with space $O(n \log \log n)$, query time $O(\log\log n)$ and polylogarithmic stretch by just repeating the above oracle-constructing algorithm for $O(\log n)$  times. Whp,
 in one of the executions the  oracle's space will be 
$O(n \log \log n)$.
Similarly, by the same Markov's argument, Corollary \ref{cor:multilevel} implies that whp one can have  the space of the oracle $\Lambda_h$ bounded by $O(n)$ (in addition to the space required to store the input graph). 

Next we analyze the construction time of our oracle.
The $h$ forests rooted at landmarks can be constructed in $\tilde{O}(m \cdot h)$ time.
We also spend $\tO(m \cdot n) = \tO(n^2 \lambda)$ time to compute all-pairs-shortest-paths (henceforth, APSP). Then
for each ball $B_{i+1}(u)$, $u \in L_i$, we store all $i$-landmarks that belong to it. 
They can be fetched from the APSP structure in $O(1)$ time per $i$-landmark.
The expected size of this data structure is $O(|{\cal P}_i|) = O({{\rho_i^2}\over {\rho_{i+1}}}) = O(n)$. Then we produce all possible quadruples $u,v,x,y$ with $v,x,y \in \Ball_{i+1}(u) \cap L_i$, $u \in L_i$. By the proof of Lemma \ref{lm:branch}, there are 
expected $O({{\rho_i^4} \over {\rho_{i+1}^3}} ) = O(n)$ such quadruples.  For each of these quadruples we check if the involved shortest paths intersect, and compute the corresponding branching events. Since the length of each such path is whp $O({n \over {\rho_{i+1}}} \cdot \log n)$, it follows that the entire computation can be carried out in $\tilde{O}({{n^2} \over {\rho_{i+1}}})$ expected  time. Recall that $\rho_{i+1} = \tilde{\Omega}(n^{1/4})$, and thus this running time is $\tilde{O}(n^{7/4})$. 
In $O(n \cdot P^2) = \tO(n^2)$ additional time we construct the DPPRO $\cL_h$ for the set of all pairs of $h$-landmarks. The total expected 
construction time is therefore dominated by the APSP computation, i.e., it is $\tO(m \cdot n)$.

\subsection{Spanner-Based Oracles}
\label{sec:sp_oracles}
While the query time of our oracle $\tLambda$ is close to optimal (there is an additive slack of $O(\log\log n)$), its space requirement $O(n \log\log n)$ is slightly suboptimal, and also its stretch requirement is $O(\log^\stretchexp n)$, instead of the desired $O(\log n)$.
Next we argue that one can get an optimal space $O(n)$ and optimal stretch $O(\log n)$, at the expense of increasing the query time to $O(n^\eps)$, for an arbitrarily small constant $\eps> 0$.

Given an $n$-vertex weighted graph $G= (V,E,\omega)$ we start with constructing an $O(\log n)$-spanner $G' = (V,H,\omega)$ of $G$ with $O(n)$ edges.
(See \cite{ADDJ90}; a faster algorithm was given in \cite{RTZ05}. For unweighted graphs a linear-time construction can be found in \cite{PS89}, and a linear-time construction with optimal stretch-space tradeoff can be found in \cite{HZ00}.) Then we build the oracle $\Lambda_h$ for the spanner $G'$. The space required by the oracle is (by Corollary \ref{cor:multilevel})  $O(n)$, plus the space required to store the spanner $G'$, i.e., also $O(n)$. Hence the total space required for this spanner-based oracle is $O(n)$.
Its stretch is the product of the stretch of the oracle, i.e., at most $t^\stretchexp$, with $t = (4/3)^h$ for an integer $h$,  and the stretch of the spanner, i.e., $O(\log n)$. Hence the oracle's stretch is $O(t^\stretchexp \cdot \log n)$. The oracle reports paths in $G' = (V,H)$, but since $H \subseteq E$, these paths belong to $G$ as well.
Observe also that the query time of the spanner-based oracle is $\tO(n^{1/t} \cdot {m' \over n})$, where $m' = |H|$ is the number of edges in the spanner. Since $m' = O(n)$, it follows that the query time is, whp, $\tO(n^{1/t})$. We remark also that the spanners produced by \cite{ADDJ90,RTZ05} have constant arboricity, and thus
one does not really need the reduction described in Section \ref{sec:extension} for this result.

\begin{theorem}
\label{thm:opt}
For any constant $\eps > 0$, the oracle obtained by invoking the oracle $\Lambda_h$ with $h = \lceil \log_{4/3} {\eps^{-1}} \rceil $ 
from Corollary \ref{cor:multilevel} on a linear-size  $O(\log n)$-spanner is a path-reporting oracle with space $O(n)$, stretch $O(\log n)$, and query time $O(n^\eps)$.

Generally, we can use an $O(k)$-spanner, ${{\log n} \over {\log\log n}} \le k \le \log n$ with $O(n^{1+1/k})$ edges. As a result we obtain a path-reporting distance oracle with space $O(n^{1+1/k})$, stretch $O(k)$ and query time $O(n^{\eps + 1/k})  = O(n^{\eps+o(1)})$. 
\end{theorem}

Observe that Theorem \ref{thm:opt} exhibits an optimal (up to constant factors) tradeoff between the stretch and the oracle size
in the range  ${{\log n} \over {\log\log n}} \le k \le \log n$.
 The only known oracle that exhibits this tradeoff is due to Mendel and Naor \cite{MN06}. However, the oracle of \cite{MN06} is not path-reporting, while our oracle is.

The construction time of this oracle consists of the time required to build the $O(\log n)$-spanner (which is $\tilde{O}(n^2)$  \cite{RTZ05}) and the construction time of the oracle $\Lambda_h$ in $G'$ (which is also $\tilde{O}(n^2)$, because $G'$ has $O(n)$ edges). Hence its overall construction time is $\tilde{O}(n^2)$.

In the context of unweighted graphs the same idea of invoking our oracle from Corollary \ref{cor:multilevel} on a spanner can be used in conjunction with $(1+\eps,\beta)$-spanners. Given an unweighted $n$-vertex graph $G= (V,E)$, let $G' = (V,H)$ be its $(1+\delta,\beta)$-spanner, $\beta = \beta(\delta,k) = \left({{\log k} \over \delta}\right)^{O(\log k)}$, with $|H| = O(\beta \cdot n^{1+1/k})$ edges, for a pair of parameters $\delta > 0$, $k=1,2,\ldots$. 
(Such a construction was devised in \cite{EP01}.)
For the sake of the following application one can set $\delta=1$. 
Invoke the distance oracle from Corollary \ref{cor:multilevel} with a parameter $t$ on top of this spanner. We obtain a path-reporting distance oracle with space $O(\beta n^{1+1/k})$ (whp). Its stretch is $(O(t^\stretchexp),\beta = \beta(t,k))$, $\beta(t,k) = O(t^\stretchexp \cdot \beta(1,k) ) = 
t^\stretchexp \cdot k^{O(\log\log k)}$, and its query time is 
$\tO(n^{1/t + 1/k})$, whp.
As long as $t = o(k^{1\over \stretchexp})$, the multiplicative stretch is $o(k)$, the additive stretch is still $\beta(k) = k^{O(\log\log k)}$,  while the space is $O(\beta n^{1+1/k})$.
In particular, one can have query time $n^{O\left({k^{-{1 \over {\stretchexp+\eta}}}}\right)}$, for an arbitrarily small 
 constant $\eta > 0$, stretch $(o(k),k^{O(\log\log k)})$, and space $O(k^{O(\log\log k)} n^{1+1/k})$.

Another variant of this construction has a higher query time $O(n^\eps)$, for some arbitrarily small  constant $\eps > 0$, but its multiplicative stretch is $O(1)$.
We just set $t$ to be a large fixed constant and consider $k \gg t^\stretchexp$.
Then the query time is $O(n^\eps)$ whp ($\eps = t^{-1}$), stretch is $(O(1),poly(1/\eps)\cdot k^{O(\log\log k)})$, and space $O(\beta \cdot n^{1+1/k})$.

\begin{theorem}
\label{thm:EPoracle}
For any unweighted undirected $n$-vertex graph $G$, any arbitrarily small constant $\eps >0$ and any parameter $k = 1,2,\ldots$, our path-reporting distance oracle has query time $O(n^\eps)$ (whp), stretch $(O(1), \beta(k)))$ and space $O(\beta(k) \cdot n^{1+1/k})$ (whp), where $\beta(k) = k^{O(\log\log k)}$. Another variant of this oracle has query time $n^{O\left({k^{-{1 \over {\stretchexp+\eta}}}}\right)}$ whp,
for an arbitrarily small constant $\eta > 0$, stretch $(o(k),k^{O(\log\log k)})$, and space $O(k^{O(\log\log k)} \cdot n^{1+1/k})$ whp.
\end{theorem}

To our knowledge these are the first distance oracles whose tradeoff between multiplicative stretch and space is better than the classical tradeoff, i.e., $2k-1$ versus $O(n^{1+1/k})$. Naturally, we pay by having an additive stretch. By lower bounds from \cite{TZ01}, an additive stretch of $\Omega(k)$ is inevitable for such distance oracles. 

One can also use a $(5 + \eps,k^{O(1)})$-spanner with $O(n^{1+1/k})$ edges from \cite{Pet09} instead of $(1+\eps,({{\log k} \over {\eps}})^{O(\log k)})$-spanner with $({{\log k} \over {\eps}})^{O(\log k)} n^{1+1/k}$ edges from \cite{EP01} for our distance oracle. As a result the oracle's space bound decreases to $O(n^{1+1/k})$, its additive stretch becomes polynomial in $k$, but the multiplicative stretch grows by a factor of $5+\eps$.
In  general, any construction of $(\alpha,\beta)$-spanners with size $O(S \cdot n)$ can be plugged in our oracle. The resulting oracle will have stretch $(t^\stretchexp          \cdot \alpha, t^\stretchexp \cdot \beta)$, size $O(Sn + n \cdot \log t)$, and query time $O(S \cdot n^{1/t})$.

The construction time of this oracle is the time needed to construct the $(1+\epsilon,\beta)$-spanner $G'$, plus the construction of $\Lambda_h$ on $G'$. The construction time of \cite{EP01} is $O(n^{2+1/k})$. The construction time of the oracle $\Lambda_h$ on $G'$ is
 $\tO(m' \cdot n')$, where $m' = O(\beta \cdot n^{1+1/k})$ is the number of edges in $G'$, and $n' = n$ is the number of vertices in $G'$. Hence the overall  construction time in this case is $O(\beta(k) \cdot n^{2+1/k}) = k^{O(\log\log k)} n^{2+1/k}$.

\section{Lower Bounds}
\label{sec:lb}

In this section we argue that one cannot expect to obtain distance labeling or routing schemes (see Section \ref{sec:prel} for their definitions) with properties analogous to those of our distance oracles (given by Theorem \ref{thm:EPoracle} and Corollary \ref{cor:expected}).  We also employ lower bounds of Sommer \etal \cite{SVY09} to show that a distance oracle with stretch $(O(1), \beta(k))$ and space $O(\beta(k) \cdot n^{1+1/k})$ for unweighted $n$-vertex graphs (like the distance oracle given by Theorem \ref{thm:EPoracle}) must have query time $\Omega(k)$.

\subsection{Distance Labeling and Routing}

We start with discussing distance labeling schemes. Suppose for contradiction that there were a distance labeling scheme $\cD$ for unweighted $n$-vertex graphs with maximum label size $O(n^{1 \over {t+4}})$ and stretch $(t,t \cdot \beta(k))$, for some fixed function $\beta(\cdot)$, and any parameter $k$.
Consider an infinite  family of $n$-vertex unweighted graphs $G_n = (V,E_n)$ with girth at least $t+2$ and $|E_n| = \Theta(n^{1+{1 \over {t+2}}})$. (Such a family can be easily constructed by probabilistic method; see, e.g., \cite{Bolbook98}, Theorem 3.7(a). Denser extremal graphs can be found in \cite{LPS88,LU95}.)
There are $2^{\Theta(n^{1+ {1 \over {t+2}}})}$  different subgraphs of each $G_n$. To achieve stretch $t$, one would need $2^{\Theta(n^{1+ {1 \over {t+2}}})}$   distinct encodings for these graphs, i.e., the total label size for this task is $\Omega(n^{1+{1 \over {t+2}}})$, and the maximum individual label size is $\Omega(n^{1 \over {t+2}})$. (See. e.g., \cite{TZ01}, Chapter 5, for this lower bound.)

Replace every edge of $G = G_n$ by a path of length $10t \cdot \beta(k)$, consisting of new vertices. The new graph $G'_n$ has $N = O(n^{1 + {1 \over {t+2}}}  \cdot t \cdot \beta(k))$ vertices. Invoke the distance labeling scheme $\cD$ on $G'_n$. For a pair of original vertices $u,v$ (vertices of $G_n$), the distance between them in $G'_n$ is $d'(u,v) = 10t \beta(k) \cdot d_G(u,v)$. Given their labels $\varphi(u)$ and $\varphi(v)$, the labeling scheme $\cD$ provides us with an estimate $\delta(\varphi(u),\varphi(v))$ of the distance between them in $G'_n$ which satisfies: 
$$\delta(\varphi(u),\varphi(v))  ~\le ~ t \cdot d'(u,v) + t \cdot \beta(k) = (10 t \beta(k) \cdot d_G(u,v))\cdot t + t \cdot \beta(k)~.$$
On the other hand, a path of length $d_G(u,v) \cdot t + 1$ in $G$ between $u$ and $v$ translates into a path of length at most 
$$10t \cdot \beta(k) (d_G(u,v) \cdot t +1) = 10 t^2 \beta(k) d_G(u,v) + 10 t \beta(k)$$
between them in $G'_n$.
Hence the estimate provided by $\cD$ corresponds to a path between $u$ and $v$ of length at most  $d_G(u,v) \cdot t$ in $G_n$, i.e., via $\cD$ we obtain a $t$-approximate distance labeling scheme for $G_n$.

The maximum label size used by $\cD$ is 
$$O(N^{1 \over {t+4}}) = O((n^{{t+3} \over {t+2}} \cdot  t \cdot \beta(k))^{1 \over {t+4}}) = O(n^{{t+3} \over {(t+2)(t+4)}} \cdot  (\beta(k))^{1 \over {t+4}})~.$$
However, by the above argument, this label size must be  $\Omega(n^{1 \over {t+2}})$.
Note that $$n^{{t+3} \over {(t+2)(t+4)}} (\beta(k))^{1 \over {t+4}} < n^{1 \over {t+2}}~,$$ as long as $\beta(k) < n$. This condition holds for any constant $k$ and fixed function $\beta(\cdot)$, and also for any $k =  O(\log n)$ and quasi-polynomial function $\beta( \cdot)$. (Recall that in all relevant upper bounds for spanners/distance oracles/distance labeling schemes, it is always the case that $k = O(\log n)$ and $\beta(\cdot)$ is at most a quasi-polynomial function of $k$.)
Hence this is a contradiction, and there can be no distance labeling scheme for unweighted graphs with label size $O(n^{1 \over {t+4}})$ and stretch $(t, t\cdot \beta(k))$, for any parameter $k$.

The same argument clearly applies to routing schemes as well. The only difference is that one needs to use lower bounds on the tradeoff between space and multiplicative stretch for routing due to \cite{PU89b,TZ01b,AGM06}, instead of analogous lower bounds of \cite{TZ01} for distance labeling.

To summarize, while Theorem \ref{thm:EPoracle} provides a distance oracle with stretch $(t,t \cdot \beta(k))$ and {\em average}  space per vertex of 
 $O(\beta(k) \cdot n^{1/k})$ for $ k \gg t^{\log_{4/3} 7}$, for distance labeling or routing one needs at least $n^{\Omega(1/t)}$   space per vertex to achieve the same stretch  guarantee. 

Similarly, one cannot have a distance labeling scheme for sparse graphs (graphs $G = (V,E)$ with $O(n^{1+1/k})$ edges, for some $k \ge 1$) with maximum label size $O(n^{1/k})$ and stretch $O(t)$, for a parameter $t \ll k$. \footnote{Recall that by Corollary \ref{cor:expected}, a path-reporting distance oracle of total size $O(n^{1+1/k})$ with stretch $O(t)$ and query time $O(n^{{1 \over {t^c}} + {1 \over k}} + |\Pi(u,v)|)$ (for a query $u,v$; the constant $c$ is given by $c = \log_7 4/3$) does exist.}
A distance labeling scheme as above requires maximum label size of $n^{\Omega(1/t)}$, as otherwise one would get a distance labeling with stretch $(t,t \cdot \mathrm{poly}(k))$ for general graphs with maximum label size $n^{o(1/t)}$, contradiction.

\subsection{Distance Oracles}

Next we argue that in the cell-probe model of computation (cf., \cite{Milterson99}), any distance oracle with size and stretch like in Theorem \ref{thm:EPoracle}
(i.e., size $O(n^{1+1/k})$ and stretch $(O(1),\beta(k))$, for a fixed function $\beta(\cdot)$) must have query time $\Omega(k)$. We rely on the following lower bound of \cite{SVY09}.

\begin{theorem} \cite{SVY09}
\label{thm:SVY}
A distance oracle with stretch $t$ using query time $q$ requires space $\cS \ge n^{1+ {c \over {t \cdot q}}}/\log n$ in the cell-probe model with $w$-bit cells, even on unweighted undirected graphs with maximum degree at most  $(t\cdot q \cdot w)^{O(1)}$, where $t = o({{\log n} \over {\log w + \log\log n}})$, and $c$ is a positive constant. 
\end{theorem}

Suppose for a contradiction that there exists a distance oracle with stretch $(t,t \cdot \beta(k))$, for a pair of parameters $t \ll k$ and a fixed function $\beta(\cdot)$, with space at most $n^{1 + {{c/2} \over {t \cdot q}}}/\log n$ (and query time $q$) for general unweighted graphs.

Let $G = (V,E)$ be an $n$-vertex unweighted graph with maximum degree at most  $(t\cdot q \cdot w)^{O(1)}$, and let $G'$ be the graph  obtained from $G$ by replacing each edge of $G$ by  a path of length $10 t \cdot \beta(k)$.
The graph $G'$ has $N \le  (t \cdot q \cdot w)^{O(1)} \cdot \beta(k) \cdot n$ vertices, and an oracle with stretch $(t,t \cdot \beta(k))$ for $G'$ can be used also as a stretch-$t$ oracle for $G$. The size of this oracle is, by our assumption, at most 
$${{(n \cdot (t \cdot q \cdot w)^{O(1)} \cdot  \beta(k))^{1 + {{c/2} \over {t \cdot q}}}  } \over {\log N}} < {{n^{1+{{c/2} \over {t \cdot q}}}} \over {\log n}} \cdot ((t \cdot q \cdot w)^{O(1)} \beta(k))^{1 + {{c/2} \over {t\cdot q}}}~.$$
As long as $( (t \cdot q \cdot w)^{O(1)} \cdot \beta(k))^{1 + {{c/2} \over {t \cdot q}}} < n^{{c/2} \over {t \cdot q}}$, i.e., as long as
\begin{equation}
\label{eq:condition_beta}
((t \cdot q \cdot w)^{O(1)} \cdot \beta(k))^{{2 \over c} t \cdot q + 1} < n~,
\end{equation}
we have a contradiction to Theorem \ref{thm:SVY}. (As the oracle uses less than $n^{1 + {c \over {t \cdot q}}}/\log n$ space and has stretch $t$ and query time $q$.)

For $k$ being at most a mildly growing function of $n$ (specifically, $k \le \log^\zeta n$, $\zeta < 1/2$), $t = o(k)$, $q \le k$, $w = O(\log n)$, 
and $\beta(\cdot)$ being a polynomial (or even a quasi-polynomial) function, the condition (\ref{eq:condition_beta}) holds.
Hence in this range of parameters, any distance oracle for unweighted graphs with stretch $(t,t \cdot \beta(k))$ and query time $q$ requires space $\cS \ge n^{1 + {{c/2} \over {t \cdot q}}}/\log n$ in the cell-probe model with $w$-bit cells, assuming $t = o({{\log n} \over {\log w + \log\log n}})$. 

So if this oracle uses $\cS = O(n^{1+1/k} \cdot \beta(k))$  space, then it holds that $n^{1+1/k} \cdot \log n \cdot \beta(k) \ge  n^{1 + {{c/2} \over {t \cdot q}}}$,
i.e.,
$$1 + 1/k + {{\log \log n + \log\beta(k)} \over {\log n}} \ge 1 + {{c/2} \over {t \cdot q}}~,$$
and so $q = \Omega(k/t)$.

We summarize this lower bound in the next theorem.

\begin{theorem}
\label{thm:do_lb}
Let $k \le \log^\zeta n$, for any constant $\zeta <  1/2$, $t =o(k)$, $w = O(\log n)$, and $\beta(\cdot)$ being a polynomial or a quasi-polynomial function.
In the cell-probe model with $w$-bit cells any distance oracle for general unweighted undirected $n$-vertex graphs with space $O(\beta(k) \cdot n^{1+1/k})$ and stretch $(t,t \cdot \beta(k))$ has query time $q = \Omega(k/t) = \Omega(k)$. 
\end{theorem}

Theorem \ref{thm:do_lb} states that in contrast to distance oracles with multiplicative stretch which can have constant query time (see \cite{MN06,C14}), a distance oracle with stretch $(O(1),\beta(k))$ (like the one given by our Theorem \ref{thm:EPoracle}) must have query time $\Omega(k)$.

\section*{Acknowledgements}

The first-named author wishes to thank Ofer Neiman and Christian Wulff-Nilsen for helpful discussions, and Elad Verbin for explaining him the lower bounds from \cite{SVY09}.

\bibliographystyle{abbrv}{
\bibliography{oracle}  
}

\clearpage
\pagenumbering{roman}
\appendix      
\centerline{\LARGE\bf Appendix}

\comment
\section{An Extension to General Graphs}
\label{sec:extension}

In this section we argue that Theorem \ref{thm:basic} can be extended to general $n$-vertex graphs $G = (V,E,\omega)$ with $m = \lambda n$ edges.
In its current form the theorem only applies to graphs of arboricity at most $\lambda$. While this is sufficient for our main application, i.e., for Theorem \ref{thm:opt}, our another application (Theorem \ref{thm:EPoracle}) requires a more general result. We remark that our extension is based on the reduction of Agarwal \etal \cite{AGHP11} of the distance oracle problem in general graphs to the same problem in bounded-degree graphs. Our reduction is somewhat more general than the one from \cite{AGHP11}, as it also applies to path-reporting distance oracles. We provide our extension for the sake of completeness.

Given an $m$-edge $n$-vertex graph $G$ with $\lambda = m/n$, we split each vertex $u_i$ into $d(u) = \lceil {{\deg(u)} \over \lambda} \rceil$ copies $u^{(1)},u^{(2)},\ldots,u^{(d(u))}$. Each copy is now selected independently at random with probability $\rho/n$, for a parameter $\rho$ determined in the same way as in Section \ref{sec:basic}. The original vertex $u$ is selected to the landmarks' set if and only if at least one of its copies (which will also be called {\em virtual nodes}) is selected. Observe that the rule that we have described is equivalent to selecting $u$ with probability $d(u) \cdot {\rho \over n} = \lceil {{\deg(u)} \over \lambda } \rceil \cdot {\rho \over n}$.

The expected number of selected virtual nodes is 
$$\sum_{v \in V} d(v) \cdot {\rho \over n} ~=~ {\rho \over n} \cdot \sum_{v \in V} \lceil {{\deg(u)} \over \lambda} \rceil~ \le~ {\rho \over n}  \sum_{v \in V} ({{\deg(v)} \over \lambda} +1)  ~=~ \rho + {{\rho} \over {\lambda n}}\sum_{v \in V} \deg(v) ~=~ 3\rho~.$$
The number $|L|$ of landmarks is at most the number of selected virtual nodes, and so $\Expect(|L|) \le 3\rho$. By Chernoff's bound, the number of selected virtual nodes is whp $O(\rho)$, and so, whp, $|L|^{2+2/k} = O(\rho^{2+2/k})$ as well. Hence the size of our oracle remains $O(n)$.

The rest of the construction algorithm for our distance oracle is identical to that of Section \ref{sec:basic}. (The only change is the distribution of selecting landmarks.) 
The query algorithm is identical to the query algorithm from Section \ref{sec:basic}. In particular, note that the virtual nodes have no effect on the computation, i.e., the returned paths contain only original vertices.

Next we argue that the expected query time of the modified oracle is still at most $O({n \over \rho} \cdot \lambda)$ in unweighted graphs, and $O({n \over \rho} \cdot \lambda \log n)$ in weighted ones. (As usual, we omit the additive term of the number of edges of the returned path.)
Specifically, we argue that the tests if $v \in Ball(u)$ and if $u \in Ball(v)$ can be carried out within the above expected time.

Let  $u = u_0,u_1,\ldots,u_{n-1}$ be all graph vertices ordered by a Dijkstra exploration originated from $u$, and replace each vertex $u_i$ by its $d(u_i)$ copies $u_i^{(1)},\ldots,u_i^{(d(u_i)}$, The copies appear in an arbitrary order. 
Since each virtual node has probability ${\rho \over n}$ to be selected independently of other vertices, it follows by a previous argument that the expected number $N$ of virtual nodes that the algorithm encounters before seeing a selected virtual node is $O({n \over \rho})$. (The algorithm actually explores only original vertices. For the sake of this argument we imagine that when the algorithm reaches a vertex $y$ it reaches its first copy $y^{(1)}$. Right after that it reaches the next copy $y^{(2)}$, etc., and then reaches $y^{(d(y))}$. After "reaching" all these copies the algorithm continues to the next original vertex.) 

Denote the original vertices explored by the algorithm $u_1,u_2,\ldots,u_{i-1},u_i$, and let $u_i^h$ be a selected copy of $u_i$. (We assume that all copies of $u_j$, for $j < i$, are not selected, and all copies $u_i^{h'}$, $h'  < h$, are also not selected.) It follows that $N = \sum_{j=1}^{i-1} d(u_j) + h$.
Hence $$\Expect(\sum_{j=1}^{i-1}d(u_j)) ~\le~ \Expect(N) ~=~ O({n \over \rho})~.$$
 Hence $$\Expect( \sum_{j=1}^{i-1} \lceil {{\deg(u_j)} \over \lambda} \rceil ~=~ O({n \over \rho})$$ as well. Thus $$\Expect(\sum_{j=1}^{i-1} \deg(u_j)) ~=~ O({{\lambda n} \over \rho}) ~=~ O({m \over \rho}).$$

Observe that the number of edges explored by the algorithm before reaching $u_i$ is at most $\sum_{j=1}^{i-1} \deg(u_j)$. (The only  edges incident on $u_i$ 
explored by the algorithm are edges $(u_j,u_i)$, for $j < i$. These edges are accounted for in the above sum of degrees.) Hence the expected number of edges explored by the algorithm is $O({m \over \rho})$. Hence its expected running time is $O({m \over \rho})$ (respectively, $O({m \over \rho} \cdot \log n)$) in unweighted (resp., weighted) graphs. The bounds that hold with high probability are higher by a factor of $O(\log n)$.

\begin{corollary}
\label{cor:extended}
Up to constant factors, the result of Theorem \ref{thm:basic} holds for general $m$-edge $n$-vertex graphs with $m = \lambda n$.
\end{corollary}

This result provides a path-reporting analogue of the result of Agarwal \etal \cite{AGHP11}, which provides stretch $O(k)$ and query time $(n\lambda)^{O(1/k)}$.
 Their oracle is not path-reporting. Our oracle is path-reporting, but its query time is significantly higher, specifically it is $n^{1/2 + O(1/k)} \cdot \lambda$.
\commentend
    
\section{Missing proofs}
\label{sec:missing_proofs}

In this section we provide proofs of  Lemmas \ref{lm:quadruples} and \ref{lm:branch}.

\inline Proof of Lemma \ref{lm:quadruples}:
\PFquadlemma

\inline Proof of Lemma \ref{lm:branch}:
\PFbranchlemma

\end{document}